\documentclass{article}

\usepackage{arxiv_gs}

\usepackage[utf8]{inputenc} 
\usepackage[T1]{fontenc}    
\usepackage{hyperref}       
\usepackage{url}            
\usepackage{booktabs}       
\usepackage{amsfonts}       
\usepackage{nicefrac}       
\usepackage{microtype}      
\usepackage{lipsum}

\usepackage{natbib} 
\usepackage{lineno,hyperref}
\usepackage{lscape}
\usepackage{todonotes}
\usepackage{multirow}
\usepackage{xcolor,colortbl,rotating}
\usepackage{longtable}
\usepackage{scrextend}
\usepackage{setspace}
\usepackage{caption}
\usepackage{arydshln}
\usepackage{multibib}
\newcites{studies}{References of computational parallelization studies}
\newcites{appendix}{References of Appendix}

\usepackage{graphicx}
\DeclareGraphicsExtensions{.pdf,.png}

\title{Parallel computational optimization in operations research: A new integrative framework, literature review and research directions}
\shorttitle{Parallel computational optimization in operations research}

\author{
  Guido Schryen\thanks{I am grateful for the support provided by Abdullah Burak, Philip Empl, Constanze Hilmer, Gerhard Rauchecker, Richard Schuster, Henning Siemes, and Melih Yilmaz, who supported me substantially in searching and coding research articles.} \\
  Department of Management Information Systems\\
  Paderborn University, Germany\\
  \texttt{guido.schryen@upb.de} \\
	\texttt{www.misor.org} \\
}

\begin{document}
\maketitle

\begin{abstract}
Solving optimization problems with parallel algorithms has a long tradition in OR. Its future relevance for solving hard optimization problems in many fields, including finance, logistics, production and design, is leveraged through the increasing availability of powerful computing capabilities. Acknowledging the existence of several literature reviews on parallel optimization, we did not find reviews that cover the most recent literature on the parallelization of both exact and (meta)heuristic methods. However, in the past decade substantial advancements in parallel computing capabilities have been achieved and used by OR scholars so that an overview of modern parallel optimization in OR that accounts for these advancements is beneficial. Another issue from previous reviews results from their adoption of different foci so that concepts used to describe and structure prior literature differ. This heterogeneity is accompanied by a lack of unifying frameworks for parallel optimization across methodologies, application fields and problems, and it has finally led to an overall fragmented picture of what has been achieved and still needs to be done in parallel optimization in OR. 
This review addresses the aforementioned issues with three contributions: First, we suggest a new integrative framework of parallel computational optimization across optimization problems, algorithms and application domains. The framework integrates the perspectives of algorithmic design and  computational implementation of parallel optimization. Second, we apply the framework to synthesize prior research on parallel optimization in OR, focusing on computational studies published in the period 2008-2017. Finally, we suggest research directions for parallel optimization in OR.  
\end{abstract}


\section{Introduction}
\label{sec:introduction}

Parallel optimization has received attention in the operations research (OR) field already for decades. Drawing on algorithmic and computational parallelism in OR is appealing as real-life optimization problems in a broad range of application domains are usually NP-hard and even the implementation of (meta)heuristic optimization procedures may require substantial computing resources. It has been argued that parallelism is crucial to make at least some problem instances tractable in practice and to keep computation times at reasonable levels \citep{talbi2009metaheuristics,crainic2006}.\footnote{Impressive computational results of applying parallelization to the traveling salesman problem (TSP) are reported by \citet[p.2]{crainic2006}.} However, unsurprisingly, the application of parallel optimization has been hesitant because i) parallelizing algorithms is challenging in general from both the algorithmic and the computational perspective, and ii) a viable alternative to parallelizing algorithms has been the exploitation of ongoing increases of clock speed of single CPUs of modern microprocessors. But this growth process reached a limit already several years ago due to heat dissipation and energy consumption issues \citep{diaz2012survey}. This development makes parallelization efforts (not only in optimization) much more important than it was in earlier times.  

Fortunately, the need for parallelization has been acknowledged and accompanied by an increased availability of parallel computing resources. This availability is rooted in two phenomena: a) the rapid development of parallel hardware architectures and infrastructures, including multi-core CPUs and GPUs, local high-speed networks and massive data storage, and of libraries and software frameworks for parallel programming \citep{talbi2009metaheuristics,crainic2006,brodtkorb2013gpu}; b) the increased availability of parallel computing resources as commodity good to researchers, who have (free or low-priced) access to multi-core laptops and workstations, and even to high-performance clusters offered by universities and public cloud providers.  

The benefits of exploiting parallel processing for optimization algorithms are multi-faceted. Searching the solution space can be speeded up for both exact and (meta)heuristic algorithms so that the optimal solution or a given aspiration level of solution quality, respectively, can be achieved quicker. Implementations can also benefit from improved quality of the obtained solutions, improved robustness, and solvability of large-scale problems \cite[p. 460f]{talbi2009metaheuristics}.  

We found many published reviews on parallel optimization for particular problems, methodologies, applications, research disciplines, and technologies. Reviews of parallelization for particular optimization problems were provided for one-dimensional integer knapsack problems \citep{Gerasch1994}, vehicle routing problems (VRPs) \citep{crainic2008parallel}, non-linear optimization \citep{ISI:A1988L980500001}, mixed integer programming \citep{nwana2000parallel} and multiobjective optimization \citep{Nebro2005}. Most of the reviews that we found focus on parallel optimization regarding particular methodologies. While branch-and-bound algorithms have been reviewed by \citet{gendron1994parallel}, the majority of methodological literature reviews have focused on metaheuristics: reviews have addressed tabu search (TS) \citep{Crainic2005}, simulated annealing (SA)\citep{Aydin2005}, variable neighborhood search (VNS) \citep{Perez2005}, Greedy Randomized Adaptive Search Procedures (GRASPs) \citep{Resende2005}, swarm intelligence algorithms \citep{tan2016survey}, particle swarm optimization algorithms \citep{zhang2015comprehensive}, and different types of evolutionary algorithms, including genetic algorithms (GAs) \citep{adamidis1994review,Luque2005,cantu1998survey,alba1999survey,adamidis1994review,ISI:000281054500008}, ant colony optimization algorithms \citep{pedemonte2011survey,Janson2005}, scatter search \citep{Lopez2005} and evolutionary strategies \citep{Rudolph2005}. Several reviews have covered sets of metaheuristics \citep{cung2002strategies,eltit,Crainic2005aparallel,pardalos1995parallel,crainic2003parallel,crainic2010parallel,crainic2014designing,Crainic2018,crainic2019parallel,alba2013} and hybrid metaheuristics \citep{Cotta2005,Luna2005}. Application- and discipline-oriented reviews of parallel optimization have been provided for routing problems in logistics \citep{schulz2013gpu} and for parallel metaheuristics in the fields of telecommunications and bioinformatics \citep{Nesmachnow2005,Trelles2005,martins2006metaheuristics}. Reviews that focus on particular parallelization technologies (in particular, General Purpose Computation on Graphics Processing Unit (GPGPU)) have been proposed by \citet{boyer2013recent}, \citet{tan2016survey} and \citet{schulz2013gpu}.
 
 We acknowledge the excellent work provided in these reviews, from which our review has benefited substantially. At the same time, we see several arguments that call for a new literature review.  First, we did not find reviews that cover the most recent literature on the parallelization of both exact and (meta)heuristic methods published in the decade 2008-2017. During this time, substantial advancements in parallel computing capabilities and infrastructures have been achieved and used by  many  OR scholars so that an overview of modern parallel optimization in OR that accounts for these advancements when synthesizing and classifying the literature is beneficial. Second, based on different foci adopted in previous literature reviews, the concepts used to describe and structure prior literature differ. This heterogeneity is accompanied by a lack of unifying frameworks for describing parallel optimization across methodologies, application fields, and problems. This has led finally to an overall fragmented picture of what has been achieved and what still needs to be done in parallel optimization in OR.  As a side effect, the heterogeneity with which parallelization studies in OR have been described in terms of algorithmic parallelization, computational parallelization and performance of parallelization is high, which is beneficial from a diversity perspective but also raises problems: First, it remains unclear for authors what should be reported in an OR study that draws on parallel optimization; second, our own experience based on screening and reading several hundreds of articles is that the heterogeneity makes it often time-consuming and in some case even impossible for readers to identify the aforementioned parallelization characteristics of a study, to classify the study accordingly and to compare studies with each other.

Accounting for the aforementioned challenges, we provide three contributions in this literature review. First and to our best knowledge, we suggest the first universally applicable framework for parallel optimization in OR, which can be used by researchers to systematically describe their parallelization studies and position these in the landscape of parallel optimization without requirements on the application domain touched, the problem addressed, the methodology parallelized or the technology applied. In particular, the suggested framework integrates both algorithmic design and computational implementation issues of parallel optimization, which are usually being addressed separately in the literature. Second, we apply the integrative framework to synthesize prior research on parallel optimization in the field of OR published in the decade 2008-2017, focusing on those studies which include computational experiments. 
Finally, we suggest research directions, including recommendations, for prospective studies on parallel optimization in OR.   

We structure our review as follows: In Section \ref{sec:parallelization_framework}, we develop a framework for computational studies on parallel optimization. In Section \ref{sec:scope_selectionProcess}, we define the scope and literature selection process of our review, before we review the literature in Section \ref{sec:literature_survey} based on the suggested framework. We provide research directions for future research in Section \ref{sec:researchDirections} before we conclude our review in Section \ref{sec:conclusion}.

\section{Parallelization Framework}
\label{sec:parallelization_framework}

Computational studies on parallel optimization usually report on four perspectives of parallelization \citep{gendron1994parallel,Alba2005a,Crainic2005aparallel,talbi2009metaheuristics,pedemonte2011survey,Crainic2018,crainic2019parallel}: \emph{object of parallelization}, \emph{algorithmic parallelization}, \emph{computational parallelization} and \emph{performance of parallelization}.
While our review of the literature revealed that most studies make either implicitly or explicitly use of the aforementioned perspectives, we also observed a high level of heterogeneity in terms of terminology, taxonomies of parallel algorithmic design, granularity of information on parallel implementation, and performance metrics used to report computational results.
As a consequence, with an increasing body of computational studies, it has become challenging to gain an overview of computational achievements, to compare studies in terms of their achievements, to develop consistent taxonomies for computational studies, and to identify white spots that need further research. 

\begin{landscape}
\begin{figure}%
\center
\fbox{\includegraphics[width=1.5\textwidth]{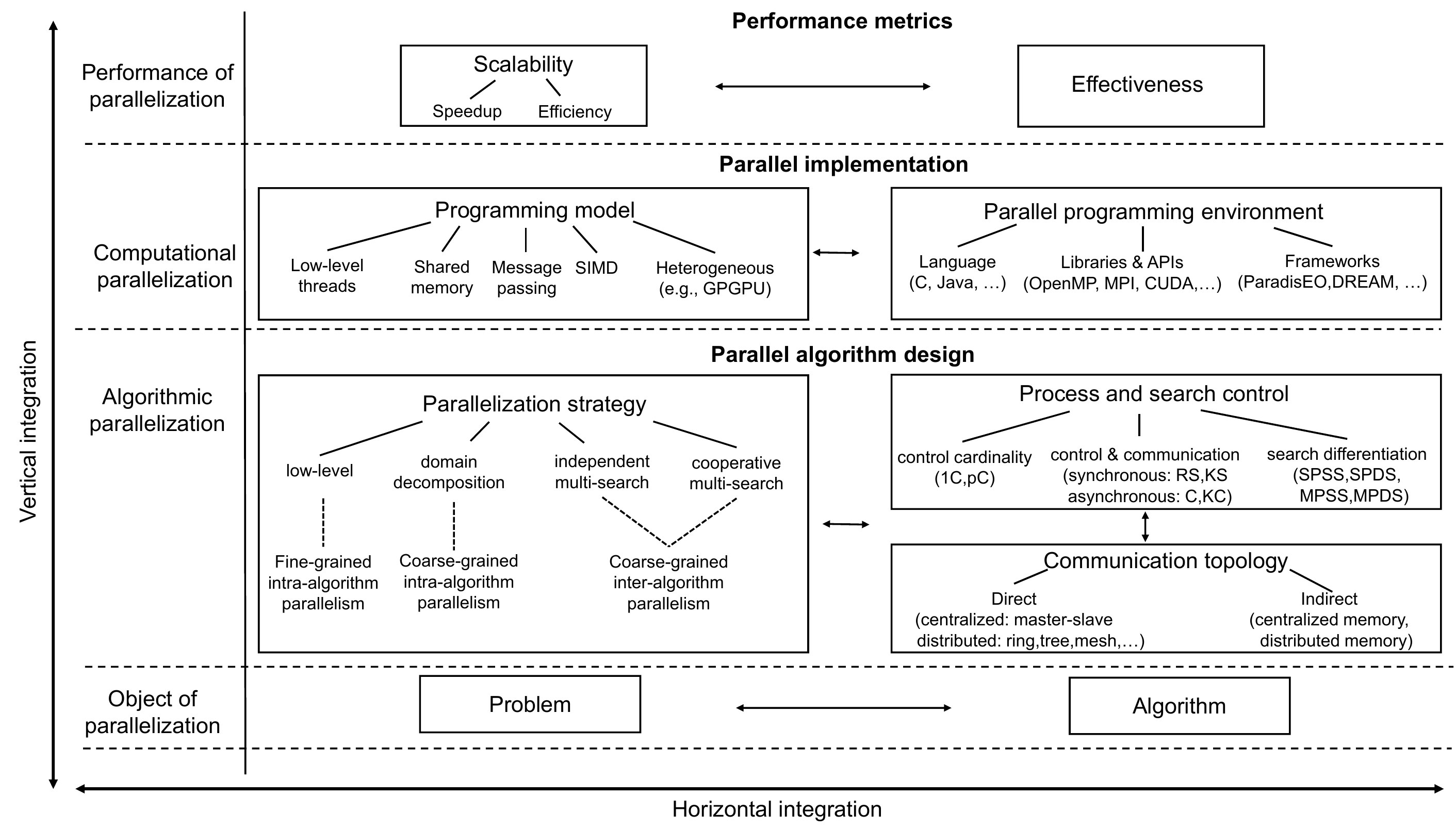}}
\caption{Integrative framework for computational parallelization studies in OR}
\label{fig:framework}
\end{figure}
\end{landscape}

In order to mitigate the aforementioned problems in the field of parallel optimization, we suggest a new descriptive framework of computational parallel optimization studies (see Figure \ref{fig:framework}). The scope of the applicability of the proposed framework in the area of parallel optimization is wide with regard to two dimensions: First, it does not make any assumptions about the addressed application domain, the optimization problem to solve, the parallelized methodology or the applied technology. We denote this broad applicability as \emph{horizontal integration}, referring to the horizontal layers in Figure \ref{fig:framework}. Second, it integrates the aforementioned perspectives (layers) and is based on well-established principles in the literature on algorithmic and computational parallelization. Similarly, we refer to this broad applicability as \emph{vertical integration}, which brings together the -- usually separately applied -- perspectives on parallel optimization found in the disciplines of OR and computer science. In this context, our framework adopts an integrated view on parallel optimization.

\subsection{Object of parallelization}
\label{sec:object_parallelization}

The object of parallelization comprises the OR problem to be solved (e.g., TSP, VRP, JSSP) and the algorithm to be applied (e.g., b\&b, GA, SA, TS), which effect each other. Problem types and algorithm types are both described in detail in Section \ref{sec:PrbTypesParAlg}. 

\subsection{Algorithmic parallelization}
\label{sec:algorithmic_parallelization}

The algorithmic parallelization refers to the methodological perspective on how parallelism is applied to solve an optimization problem by decomposition. As suggested for metaheuristics \citep{crainic2019parallel} , we detail this perspective by distinguishing various types of \emph{parallelization strategy}, \emph{process and search control}, and \emph{communication topology} (see Figure \ref{fig:framework}). Parallelization strategies have been defined according to the source of parallelism \citep{cung2002strategies,crainic2003parallel,Crainic2005aparallel,crainic2010parallel,crainic2019parallel}. Four types are distinguished: (1) Functional parallelism applies when decomposition occurs at the algorithm level by, for example, evaluating neighbor solutions or computing the fitness of a solution in parallel. This parallelization strategy does not alter the algorithmic logic, the search space or the behavior of the sequential version, and it is thus also referred to as \emph{low-level}. As parallelism occurs at a low level inside a single algorithm, we coin the term \emph{fine-grained intra-algorithm parallelism}. Since the overall search follows only a single search path, this type of parallelism has also been denoted as \emph{single-walk parallelization}, in contrast to the following strategies, where the overall search follows multiple trajectories and are referred to as \emph{multiple-walk parallelization strategies} \citep{cung2002strategies}. (2) \emph{Domain decomposition} refers to the approach of separating and exploring the search space explicitly yielding a number of smaller and easier to solve subproblems to be addressed simultaneously by applying the same sequential algorithm. The partial solutions are finally used to reconstruct an entire solution of the original problem. The separation of the search space may be obtained, for example, by discarding or fixing variables and constraints. This separation may result in a partition (disjoint subsets) or a coverage (subsets may overlap) of the overall search space. In contrast to the low-level strategy, where parallelism occurs at a local and predefined part of the algorithm, domain decomposition involves concurrent explorations of subspaces using the same algorithm. Thus, we introduce the term \emph{coarse-grained intra-algorithm parallelism}. (3) Separating the search space can also be performed implicitly through concurrent explorations of the search space by different or differently parameterized methods. When the concurrent execution of methods does not involve any exchange of information prior to identifying the best overall solution at the final synchronization step, the parallelization strategy is referred to as \emph{independent multi-search}, which can be perceived as \emph{coarse-grained inter-algorithm parallelism}. (4) When the concurrent execution of methods and their explorations of subspaces involves the exchange of information through cooperation mechanisms while the search process is in progress, \emph{cooperative multi-search} occurs. The sharing of information may even be accompanied with the creation of new information out of exchanged data. As the interactions of the cooperative search algorithms specify the global search behavior, a new metaheuristic in its own right emerges \citep{crainic2008}. While cooperation yields in many cases a collective output with better solutions than a parallel independent search \citep{crainic2019parallel}, exchanges should not be too frequent to avoid communication overheads and premature ``convergence" to local optima \citep{toulouse2000global,toulouse2004systemic}. As in the case of independent multi-search, also cooperative multi-search can be seen as coarse-grained inter-algorithm parallelism. Finally, it should be noticed that parallelization strategies are not mutually incompatible and may be combined into comprehensive algorithmic designs \citep{crainic2006,crainic2019parallel}. For example, low-level and decomposition parallelism have been jointly applied to branch-and-bound \citep{Adel2016} and dynamic programming \citep{ISI:000368652500007},  \citep{ISI:000385623100022}, and low-level parallelism and cooperative multi-search have been applied to a hybrid metaheuristic \citep{ISI:000271404200004} which uses a genetic algorithm and hill climbing. 

While the aforementioned parallelization strategies have been formulated for the class of metaheuristics, the strategy-defining principles are of general nature of parallelizing optimization algorithms so that the scope of applicability of the parallelization strategies can be straightforward extended to other algorithm classes, including exact methods and (problem-specific) heuristics. For example, \citet{gendron1994parallel} have defined three types of parallelism for branch-and-bound: their type 1 parallelism refers to parallelism when performing operations on generated subproblems, such as executing the bounding operation in parallel for each subproblem. This type can be perceived as low-level parallelism. Parallelism of type 2 consists of building the branch-and-bound  tree in parallel by performing operations on several subproblems concurrently. This type of parallelism involves an explicit separation of the search space and can, thus, be perceived as domain decomposition. Finally, the case of type 3 parallelism implies that several branch-and-bound trees are built in parallel, with the trees being characterized by different operations
(branching, bounding, testing for elimination, or selection). This parallelism includes the option to use the information generated during the construction of a tree for the construction of another one. When such information is exchanged, type 3 parallelism can be perceived as cooperative multi-search, otherwise it corresponds to independent multi-search. The straightforward matching of parallelization strategies for metaheuristics with types of parallelism defined for an exact method supports our previous argument that the four parallelization strategies can be applied to the general ``universe" of optimization algorithms.   

Process and search control refers to how the global problems-solving process is controlled, how concurrent processes communicate with each other, and how diverse the overall search process is. We adopt the three dimensions suggested by \citet{Crainic2005aparallel}: \emph{Search control cardinality} determines whether the global search is controlled by a single process (1-control, 1C)) or by several processes (p-control, pC) which may collaborate or not. \emph{Search control and communications} refers to how information is exchanged between processes and distinguishes between synchronous and asynchronous communication. In the former
case, all concerned processes have to stop and engage in some form of communication and information exchange at specified moments (e.g., number of iterations) exogenously determined. In the latter case, processes are in charge of their own search as well as of establishing communications with other processes, and the global search terminates once each individual search stops. Both synchronous and asynchronous communication can be further qualified with regard to whether additional knowledge is derived from communication, leading to four categories of control and communication: rigid (RS) and knowledge synchronization (KS) in the synchronous case, and collegial (C) and knowledge collegial (KC) in the asynchronous case. Finally, the diversity of search may vary according to whether concurrently executed methods start from the same or different solutions, and to whether their search follows the same or different logics\footnote{Two logics are characterized as ``different" even when based on the same methodology (e.g., two tabu searches or genetic algorithms) if they vary in terms of components (e.g., neighborhoods or selection mechanism) or parameter values \citep{crainic2019parallel}.}; the diversity of search is also referred to as \emph{search differentiation}. From these two dimensions the following four classes can be derived: 1. \emph{same initial point/population, same search strategy (SPSS)}; 2.\emph{same initial point/population, different search strategies (SPDS)}; 3. \emph{multiple initial points/populations, same search strategies (MPSS)}; 4. \emph{multiple initial points/populations, different search strategies (MPDS)}. While the term ``point" relates to single-solution methods, the notion ``population" is used for population-based ones, such as genetic algorithms or ant colony optimizations. As in the case of parallelization strategies described above, the three dimensions of process and search control have been suggested for the classification of metaheuristics \citep{Crainic2005aparallel,Crainic2018,crainic2019parallel} but can be extended straightforward to other classes of optimization algorithms. 

When concurrent processes exchange information, they may communicate with each other in a direct or indirect way. Direct communication involves message-based communication along some communication topology, such as a tree, ring, or fully connected mesh \citep{talbi2009metaheuristics, crainic2019parallel}. This communication topology needs to be projected on a physical interconnection topology as part of the implementation design. In contrast, indirect communication involves the use of a centralized or distributed memory, which are used as shared data resources of concurrent processes \citep{crainic2019parallel}. 

The three perspectives of parallel algorithm design, namely parallelization strategy, process and search control, and communication topology, are linked together \citep{Crainic2018,crainic2019parallel}. Low-level parallelization is generally targeted in 1C/RS/SPSS designs, with the 1C (control cardinality) being implemented with a master-slave approach. Examples are the neighborhood evaluation of a local search heuristic, and the application of operators and the determination of fitness values in a GA. Domain decomposition is often implemented using a master-slave 1C/RS scheme with MPSS or MPDS search differentiation but can also be performed in a pC, collegial decision making framework with MPSS or MPDS search differentiation. Independent multi-search is inherently a pC parallelization strategy, which follows from the same or different starting point(s)/population(s) with or without different search strategies (i.e., SPDS, MPSS or MPDS search differentiation). As the concurrently executed search processes do not exchange information prior to the final step, they follow the RS control and communication paradigm. Finally, cooperative multi-search is also a pC parallelization strategy, which may start from possibly different starting points/populations and may follow different search strategies (i.e., SPDS, MPSS or MPDS search differentiation). In contrast to independent multi-search, information is exchanged between processes during the search. This exchange of information can vary in different ways, which results in a large diversity of cooperation mechanisms. First, different types of information may be exchanged, including ``good" solutions and context information. Second, cooperating processes may exchange information directly by sending messages to each other based on a given communication topology, or indirectly using memories which act as data pools shared by processes. A third option distinguishes between synchronous and asynchronous cooperation, where processes either need to stop its activities' until all others are ready or not, respectively.

\subsection{Computational parallelization}
\label{sec:computational_parallelization}

When parallel algorithms are implemented and executed in modern computational environments, different parallel programming models may be applied in a variety of programming environments. Albeit being intertwined (see, for example, \citep{talbi2009metaheuristics}), they represent different facets of parallel implementation from a conceptual perspective. Four (pure) parallel programming models can be distinguished: threads, shared memory, message passing \citep{diaz2012survey,talbi2009metaheuristics} and single-instruction-multiple-data (SIMD). In the thread programming model, lightweight processes (threads) are executed, where the communication between threads is based on shared addresses.
The shared memory programming model, where, too, tasks share a common address space, operates at a higher abstraction level than threads. Today, both the thread and the shared memory model are executed on a multi-core CPU architecture on a single computer node. In contrast, in the message passing programming model the communication between processes is done by sending and receiving messages. Each process has its own address space that is not shared with other processes. This model is designed for execution in computer clusters, where different nodes are connected through high-speed networks. Note that, depending on the particular parallel programming model, parallel executed software parts are labeled differently usually as \emph{threads}, \emph{tasks} or \emph{processes}. Finally, SIMD exploits data parallelism by operating a single instruction on multiple data on a vector processor or array processor.
Beyond the pure parallel programming models sketched above, the heterogeneous model \emph{General Purpose Computation on Graphics Processing Unit (GPGPU)} has received increasing attention (e.g., \citep{brodtkorb2013gpu}). GPGPU harnesses the capabilities of multi-core CPUs and many-core GPUs, where threads are executed in parallel on GPU cores and where GPUs can have different levels of shared memory; in this sense, we can speak of heterogeneous systems \citep{diaz2012survey}. 
Other heterogeneous models are distributed shared memory models and field programmable gate arrays (FPGAs).
In modern computing environments, (pure or heterogeneous) parallel programming models are sometimes combined with each other by, e.g., jointly using threads and GPGPU, shared memory and message passing, or threads and message passing \citep{diaz2012survey}. Such approaches are referred to as \emph{hybrid models}.

Parallel programming environments are related to parallel programming models and comprise languages, libraries, APIs (application programming interfaces) and frameworks.

\subsection{Parallel performance metrics}
\label{sec:parallel_performance_metrics}

The general purpose of parallel computation is to take advantage of increased processing power to solve problems faster or to achieve better solutions. The former goal is a matter of \emph{scalability}, which is defined as the degree to which it is capable of efficiently utilizing increased computing resources. Performance measures of scalability fall into two main groups: \emph{speedup} and \emph{efficiency}. 
Speedup $S_p:=\frac{S}{T_P}$ is defined as the ratio of sequential computation time $S$ to parallel computation time $T_p$ when the parallel algorithm is executed on $p$ processing units (e.g., cores in a multicore processor architecture). The serial time $S$ can be measured differently, leading to different interpretations of speedup \citep{barr1993reporting}: When $S$ refers to the fastest serial time on any serial computer, speedup is denoted as \emph{absolute}. Alternatively, $S$ may also refer to the time required to solve a problem with the parallel code on one processor. This type of speedup is qualified as \emph{relative}. When real-time reduction is considered as the primary objective of parallel processing, absolute speedup is the relevant type. 
While speedup relates serial to parallel times, efficiency $E_p:=\frac{S_p}{p}$ relates speedup to the number of processing units used. 
With the definition of efficiency, we can qualify speedup as \emph{sublinear speedup} ($E_p<1$), \emph{linear speedup} ($E_p=1$), or \emph{superlinear speedup} ($E_p>1$). Sublinear speedup is often due to serial parts of a parallel algorithm and several reasons for a nonvanishing serial part can be distinguished.
Superlinear speedup can occur, for example, when during the parallel execution of a branch-and-bound algorithm one processor finds a good bound early in the solution process and communicates it to other processors for truncation of their search domains \citep{barr1993reporting}.  Finally, it should be noticed that while the application of speedup and related efficiency concepts to algorithms which have a ``natural" serial version is straightforward, their unmodified application to multi-search algorithms, which are parallel in nature, does not make much sense as no basis of comparison is available.   

A second important performance measure in parallel optimization is the solution quality achieved through parallelization. Solution quality can be measured in various ways. When the optimal solution value or a bound of it is known, the relative gap to (the bound of) the optimal value can be determined. A second option is to relate the achieved solution quality with that obtained from sequential versions of the parallelized algorithm (relative improvement). However, this option requires that a sequential version of the parallel algorithm exists in terms of unchanged algorithmic logic and the trajectory through the search space. This is not the case, for example, when cooperative multi-search occurs, which defines a new algorithm due to cooperation. Finally, the solution quality obtained through parallelization may be compared with the quality of the best known solution obtained from any serial implementation (absolute improvement). Overall, the goal of achieving better solutions can be perceived as an issue of \emph{effectiveness}.

\section{Scope and literature selection process}
\label{sec:scope_selectionProcess}

The focus of our literature review lies on computational studies of parallel optimization, where physical or virtual parallel computing architectures have been applied to OR problems, such as TSPs, VRPs and FSSPs (flow shop scheduling problems). Due to the interdisciplinary nature of the OR field, such studies are not only found in OR outlets but also in those of many other disciplines, including management science, mathematics, engineering, natural sciences, combinations of engineering and natural sciences (such as chemical engineering), computer science, bioinformatics, material science, geology and medicine. While we include outlets of these disciplines in our search (see the succeeding subsection), we would like to stress that the focus of our review lies on studies on OR problems  and that it is beyond the scope of this review to identify and classify all articles of parallel optimization addressing problems in related fields or even across all fields (optimization in general).
Adopting this view, we exclude from our review, for example, mathematical studies on parallelizing matrix computations or on conjugate gradient methods, computer science studies on load balancing issues in parallel computing environments or on solving hard problems in theoretical computer science (e.g., the subset sum problem), and parallel optimization studies across fields, such as those addressing the effects of migration in parallel evolutionary algorithms.
We also exclude works on parallel optimization when their purpose lies in designing or implementing other methodologies, such as simulation, data analysis, data mining, machine learning and artificial intelligence. We further exclude meta optimization (calibrating parameters of optimization models or methodologies). We explicitly acknowledge the importance of these areas but they deserve and need dedicated literature reviews. 
Finally, from a technological perspective, we also do not consider distributed optimization that makes use of geographically dispersed computers and allows using grids,
 which comprise networks of many, often thousands or even millions of single computers.   
This field applies programming models and parallel programming environments that differ from those used in our framework, and it would need a dedicated literature review, too.

Accounting for the previously described scope of our review, we implemented different streams of literature search. A detailed description of the literature search process is provided in the online \ref{sec:app_literature_selection:process}. 
 Although having implemented different streams of search, we admit that the application of our search procedure does not guarantee to identify all computational studies of parallel optimization in OR and that we may have overlooked studies. However, we are confident to have acquired a body of literature that is sufficiently comprehensive to draw a firm picture of computational parallelization in OR during the decade 2008-2017.    

\section{Literature survey}
\label{sec:literature_survey}

In this section, we provide a synthesis of the literature  published in the decade 2008-2017. We first offer a brief meta analysis, then we analyze the body of literature with regard to which optimization problems have been solved by which (parallelized) algorithms before we present the findings of our literature analysis, structured along optimization algorithms and based upon the framework suggested above.
Findings on (i) effectiveness and (ii) parallel programming environments are not presented here because (i) effectiveness results have been reported only rarely and in  partially inconsistent ways in the studies of our sample, making comparisons of results difficult, and (ii) parallel programming environments should be considered across algorithms. We discuss both topics in Section \ref{sec:researchDirections}.
With regard to speedup, we qualify it by efficiency when reported in a study. When GPGPU is used as programming model, we only report speedup values without providing the number of parallel processing units or information on efficiency. The reason is that the number of parallel working units (usually GPGPU threads) needs to be interpreted different from that counting other parallel working units (CPU threads, processes) so that efficiency usually being defined as the ratio of speedup and the number of parallel processing units  is not applied here. Details on this issue as well as the coding of all studies in our sample are provided in the online \ref{sec:coding_parall_studies}.

\subsection{Meta analysis}

Overall, our sample consists of 206 studies, with 164 studies published in 77 different journals, 38 studies published at 36 different workshops, symposiums, conferences or congresses, and four studies published as book chapters.
The joint distribution of articles over scientific outlets and years is summarized in Table \ref{tab:outlet_year}, which shows that (1) there is no clear temporal development of the numbers of papers published per year, (2) while the number of scholarly outlets (journals, proceedings, etc.) which have published computational studies on parallel optimization in OR is high, only nine outlets have published at least five articles during the decade 2008-2017 and only three outlets (namely, \emph{Computers \& Operations Research}, \emph{European Journal of Operational Research}, \emph{Journal of Parallel and Distributed Computing}) have published more than ten articles in the same period. Overall, this publication landscape does not reveal clear clusters in terms of time or outlet, it rather shows that computational and parallel optimization in OR has been covered permanently (and) distributed over many outlets rooted in different yet related academic disciplines, including \emph{OR}, \emph{Computer Science} and \emph{Engineering}. Apparently, this research area is of multidisciplinary relevance.

\begin{table}%
\centering

\begin{tabular}{|c|*{10}{c}|c|}
  \hline
	\multirow{2}{*}{Outlet} & \multicolumn{10}{c|}{Year} & \multirow{2}{*}{Sum}\\
	\cline{2-11}
	& 2008 & 2009 & 2010 & 2011 & 2012 & 2013 &2014 & 2015 &2016 & 2017 &\\
	\hline
	ASC & & & & 2 & 1 & & 1 & 1 & 1 & 2 & 8\\
	CIE & & & 1 & & & 1 & & 1 & 1 & 1 & 5\\
	COR & 1 & & 1 & & 3 & 2 & 2 & & & 2 & 11\\
	CCPE & & & & & & 1 & & 1 & & 3 & 5\\
	EJOR & & 3 & 1 & & 1 & 1 & 1 & 3 & 2 & 2 & 13\\
	IJOC & 1 & 1 & & 2 & & & & & 1 & 1 & 6\\
	JPDC & 1 & 1 & 1 & 2 & & 4 & 1 & 1 & 1 & & 12\\
	JSC & & & & 1 & & & 1 & & 1 & 2 & 5\\
	PC & 1 & & & 1 & 1 & & & 2 & 2 & 1 & 8\\
	Other journals & 7 & 5 & 5 & 12 & 12 & 11 & 7 & 10 & 12 & 13 & 91\\	
  Proceedings& 3 & 5 & 7 & 11 & 4 & 4 & 3 & & 1 & & 37\\	
	Book chapters& 2 & 1 & & 1 & 1 & & & & & & 5\\	
	\hline
	Sum & 15 & 15 & 16 & 32 & 22 & 24 & 16 & 18 & 21 & 27 & 206\\
	\hline
	\multicolumn{12}{|l|}{ASC: Applied Soft Computing}\\
	\multicolumn{12}{|l|}{CIE: Computers \& Industrial Engineering}\\
\multicolumn{12}{|l|}{COR: Computers \& Operations Research}\\
\multicolumn{12}{|l|}{CCPE: Concurrency and Computation-Practice \& Experience}\\
\multicolumn{12}{|l|}{EJOR: European Journal of Operational Research}\\
\multicolumn{12}{|l|}{IJOC: INFORMS Journal on Computing}\\
\multicolumn{12}{|l|}{JPDC: Journal of Parallel and Distributed Computing}\\
\multicolumn{12}{|l|}{JSC: Journal of Supercomputing}\\
\multicolumn{12}{|l|}{PC: Parallel Computing}\\\hline
\end{tabular}

\caption{Joint distribution of selected articles over scientific outlets and years}
\label{tab:outlet_year}
\end{table}

\subsection{Problem types and parallelized algorithms}
\label{sec:PrbTypesParAlg}

\begin{table}%
\centering

\begin{tabular}{|c|*{13}{c}|c|} 
  \hline
	{Alg.} & \multicolumn{13}{c|}{Problem type} & \multirow{2}{*}{Sum}\\
	\cline{2-14}
	type & AP & FLP & FSSP & GTP & JSSP & KP & BFP & MILP & MSP & SOP & TSP & VRP & Other &\\
	\hline
	B-a-X & 1 & & 7 & 3 & 2 & 3 & 2 & 4 & & 2 & 3 & & 13 & 40\\
	DP & & & & 2 & & 3 & & & & 1 & & & 4 & 10\\
	IPM & & & & & & & & & & 2 & & & 2 & 4\\
	PSEA & & & & 2 & & 1 & & 1 & & & & & & 4\\
	\hline
	PSH & & 1 & 1 & 1 & & & & 1 & & & 2 & & 6 & 12\\
	\hline
  TS & 4 & & 5 & & 2 & & 1 & & 1 & &2 & 5 & 3 & 23\\
	SA & & & 2 & & 2 & & 1 & & 1 & & 1 & 3 & 1 & 11\\
	VNS & & & 1 & & 2 & 1 & & & 1 & & & 4 & 2 & 11\\
	GRAS & & & & & & & & & & & & & 2 & 2\\
	OSSH & & & & & & 1  & & & & & & & & 1\\
	\hline
	GA & 2 & 2 & 3 & 1 & 1 & & 3 & & 3 & & 3 & & 10 & 28\\
	OEA & & 1 & & & & 1 & 2 & & 1 & & 1 & 1 & 6 & 13 \\
	SSPR & & & 1 & & & & & & & & & & 1 & 2\\
	ACO & 2 & & & & & & & & & & 12 & 2 & & 16\\
	PSO & & 1 & & 2 & 1 & & 3 & & & & & & 5 & 12\\
	BCO & & & & & & & 2 & & 1 & & & & & 3\\
	FA & & & & & & & 1 & & & & & & & 1\\
	\hline
	HM & 1 & 1 & 2 & 1 & 1 & 1 & 2 & & 3 & & 4 & 2 & 7 & 25\\
	\hline
	OH & & & & 1 & & & & & & & & 1 & 2 & 4\\
	\hline
	MH & & 1 & & & & & & & & & & 1 & & 2\\
	\hline
	MS & 1 & & & & & & & & & & & 1 & 1 & 3\\
	\hline
	Sum & 11 & 7 & 22 & 13 & 11 & 11 & 17 & 6 & 11 & 5 & 28 & 20 & 65 & 227\\
	\hline
	\multicolumn{8}{|c|}{Optimization Problem Type} & \multicolumn{7}{c|}{Algorithm type}\\
	\hline
	\multicolumn{8}{|l|}{AP: Assignment Problem} & \multicolumn{7}{l|}{Exact algorithms:}\\
	\multicolumn{8}{|l|}{FLP: Facility Location Problem} & & \multicolumn{6}{l|}{B-a-X: Branch \& X}\\
	\multicolumn{8}{|l|}{FSSP: Flow Shop Scheduling Problem} & & \multicolumn{6}{l|}{DP: Dynamic programming}\\
	\multicolumn{8}{|l|}{GTP: Graph Theory Problem} & & \multicolumn{6}{l|}{IPM: Interior point method}\\	
	\multicolumn{8}{|l|}{JSSP: Job Shop Scheduling Problem} & & \multicolumn{6}{l|}{PSEA: Problem-specific exact algorithms}\\
	\cline{9-15}
	\multicolumn{8}{|l|}{KP: Knapsack Problem} & \multicolumn{7}{l|}{PSH: Problem-specific heuristics}\\
	\cline{9-15}
	\multicolumn{8}{|l|}{BFP: Benchmark function optimization problem(s)} & \multicolumn{7}{l|}{Single-solution based metaheuristics:}\\	
  \multicolumn{8}{|l|}{MILP: (Mixed) Integer Linear Program} & & \multicolumn{6}{l|}{TS: Tabu search}\\
	\multicolumn{8}{|l|}{MSP: Machine Scheduling Problem} & & \multicolumn{6}{l|}{SA: Simulated annealing}\\	
	\multicolumn{8}{|l|}{SOP: Stochastic Optimization Problem} & & \multicolumn{6}{l|}{VNS: Variable neigborhood search}\\
	\multicolumn{8}{|l|}{TSP: Traveling Salesman Problem} & & \multicolumn{6}{l|}{GRAS: (Greedy randomized) adaptive search}\\
	\multicolumn{8}{|l|}{VRP: Vehicle Routing Problem} & & \multicolumn{6}{l|}{OSSH: Other single solution heuristics}\\
	\cline{9-15}	
	\multicolumn{8}{|l|}{} & \multicolumn{7}{l|}{Population-based metaheuristics:}\\
	\multicolumn{8}{|l|}{} & &\multicolumn{6}{l|}{GA: Genetic algorithm}\\
  \multicolumn{8}{|l|}{} & &\multicolumn{6}{l|}{OEA: Other evolutionary algorithms}\\
  \multicolumn{8}{|l|}{} & &\multicolumn{6}{l|}{SSPR: Scatter search \& path relinking}\\
	 \multicolumn{8}{|l|}{} & &\multicolumn{6}{l|}{ACO: Ant colony optimization}\\
  \multicolumn{8}{|l|}{} & &\multicolumn{6}{l|}{PSO: Particle swarm optimization}\\
  \multicolumn{8}{|l|}{} & &\multicolumn{6}{l|}{BCO: Bee colony optimization}\\
  \multicolumn{8}{|l|}{} & &\multicolumn{6}{l|}{FA: Fireworks algorithm}\\
	\cline{9-15}
  \multicolumn{8}{|l|}{} & \multicolumn{7}{l|}{HM: Hybrid metaheuristics}\\
	\cline{9-15}
  \multicolumn{8}{|l|}{} & \multicolumn{7}{l|}{OH: Other heuristics}\\
	\cline{9-15}
  \multicolumn{8}{|l|}{} & \multicolumn{7}{l|}{MH: Matheuristics}\\	
	\cline{9-15}
  \multicolumn{8}{|l|}{} & \multicolumn{7}{l|}{MS: Multi-search algorithms}\\	
	\hline	
\end{tabular}

\caption{Joint distribution of selected articles over problems and (parallelized) algorithms}
\label{tab:problem_algorithm}
\end{table}     

We now describe the identified body of literature from the perspective of problem types and types of parallelized algorithms. Table \ref{tab:problem_algorithm} shows the joint distribution of articles over these two dimensions. We identified problem types by, firstly, coding for each article of our sample the covered problem(s) and, secondly, consolidating problems to problem types widely used in the OR literature\footnote{An example of consolidation is grouping the ``multi-depot VRP" and the ``VRPs with time windows" to the problem type ``VRP".} Overall, we identified nine ``application-oriented" problem types (AP, FLP, FSSP, GTP, JSSP, KP, MSP, TSP, VRP) and three ``mathematically-oriented" problem types (BFP, MILP, SOP).\footnote{While application-oriented problem types (e.g., TSP) usually lead to mathematical formulations which have an overall and coherent logic across the components (objective function, constraints, variables, etc.) of a model, ``mathematically-oriented" problem types (e.g., MILP) have mathematical formulations where single components have to meet mathematical assumptions (e.g., binary variables, linear terms) without requiring the overall model to refer to a specific application concept.} Adopting this distinction leads to assigning a study that, for example, formulates a TSP as a mixed-integer linear program to the problem class ``TSP" rather than to the class ``MILP" as it is TSP instances that are focused and not MILP instances in general. Conversely, studies assigned to one of the classes BFP, MILP or SOP explicitly address the related mathematically-oriented problem type and are not necessarily linked to any specific application
. We consolidated all problem types for which only very few computational parallelization studies have been published to the category ``Other"\footnote{When an article studies several ``other" problem types, we did not count the number of other problem types but coded it as a single appearance of an ``other problem type".}.

With regard to types of algorithms, we draw on a taxonomy suggested by \citet{talbi2009metaheuristics}, who distinguishes between \emph{exact algorithms} (e.g., branch-and-bound), \emph{problem-specific heuristics} (e.g., Lin-Kernighan heuristic for the TSP), \emph{single-solution based metaheuristics} (e.g., tabu search), and \emph{population-based metaheuristics} (e.g., genetic algorithms)\footnote{The authors also suggest the type \emph{approximation algorithms}, which we do not use in this review.}. We extend the taxonomy by adding some algorithm types: \emph{hybrid metaheuristics} refer to an metaheuristic where parts of a (meta)heuristic $A$ are embedded into a step of a (meta)heuristic $B$; \emph{matheuristics} refer to the interoperation of metaheuristics and (exact) mathematical programming techniques; \emph{multi-search algorithms} refers to the combination of several independent search algorithms, which may collaborate or not. Finally, we provide \emph{other heuristics} as a residual type for those (meta)heuristics which do not fit to any of the aforementioned algorithm types.  

It should be noticed that the sums of addressed problem types and parallelized algorithm types shown in Table \ref{tab:problem_algorithm} do not equal the sample size for different reasons: (i) some articles in our sample apply more than one algorithm type to a single problem type and/or investigate more than one optimization problem type; (ii)  
a few articles do not clearly reveal (from our perspective) the targeted problem or the applied algorithm, or they do not parallelize any algorithm but only the evaluation of the objective function; due to these reasons, we excluded five articles from the presentation in Table \ref{tab:problem_algorithm}. Overall, it should be kept in mind that each combination of addressed problem type and parallelized algorithm type is a ``case" of a study, where a single study may have several cases.       
The perspective on optimization problems addressed in computational parallelization studies shows that a broad range of problem types have been covered. Beyond the 12 problem types highlighted, the residual class of other problem types includes 63 cases, in which computational parallelization has been applied to mostly different problem types. However, we also notice that a set of 
12 problem types account for more than 70\% of all cases, with a focus on the TSP, the FSSP and the VRP, which jointly account for more than 30\% of all cases.
Similar results are obtained from adopting the algorithmic perspective. While a broad range of exact algorithms and single-solution, population-based and hybrid metaheuristics
have been parallelized, only a few algorithm types (branch-and-X (X=bound, cut, price, etc.), GAs, hybrid metaheuristics, TS) account for more than 50\% of all cases, with branch-and-X accounting for about 18\%. 
Jointly adopting the problem and algorithmic perspective, again, shows a large diversity but in this case no large clusters occur. Only four combinations (ant colony optimization applied to the TSP, branch-and-X applied to the FSSP, TS applied to the FSSP, TS applied to the VRP) have been covered in at least five cases, but these four combinations account for only about 13\% of all cases.

\begin{table}
\begin{addmargin}{-7mm}
\centering
\small

\begin{tabular}{|p{0.1cm}p{2.5cm}|p{13.0cm}|}
	\hline
	\multicolumn{2}{|l|}{Algorithm type} & Computational studies\\
	\hline
		\multicolumn{2}{|l|}{Exact algorithms:} & \\
  \cline{2-3}
		& {Branch \& X} & \citep{mezmaz2014multi,ISI:000318042500009,Herrera2017,ISI:000255647700028,ISI:000404015200003,ISI:000332452400025,ISI:000299550000011,ISI:000321416800013,ISI:000354949200009,eckstein2015pebbl,ISI:000335544500016,ISI:000395014100005,ISI:000398718400009,ISI:000322300300002,Bak2011a,ISI:000390638300008,ISI:000351682800004,barreto2010parallel,ISI:000368652500007,ISI:000342481400006,paulavivcius2009parallel,ISI:000258760900007,ISI:000326057700005,aitzai2013parallel,ISI:000401878000008,ISI:000289069700006,ISI:000268558800004,ISI:000392789400017,ISI:000352667900008,ISI:000319238800010,Adel2016,Borisenko2011,Boukedjar2012,Carneiro2011,Galea2011,Herrera2013,Estrada2011}\\
 \cline{2-3}
	  & {Dynamic programming} & \citep{Dias2013a,ISI:000362899100010,ISI:000385623100022,ISI:000261892000011,ISI:000278924200005,boyer2012solving,ISI:000380751900011,Kumar2011,Rashid2010,Tran2010}\\
 \cline{2-3}
	  & {Interior point method} & \citep{ISI:000417337900003,Hong2010,Lubin2012,Lucka2008}\\
 \cline{2-3}
	  & {Problem-specific exact algorithms} & \citep{ISI:000352249500002,Rossbory2013,Kollias20142400,Bozdag2008})\\
		\hline
\multicolumn{2}{|l|}{Problem-specific heuristics} &  \citep{Dobrian2011,ISI:000401878000008,Ismail2011,bozejko2009solving,ISI:000361986700006,ISI:000418216800023,ISI:000386524100014,ISI:000349592500005,ISI:000349884400003,ISI:000255671800016,ISI:000378100800023,Luo2015}.\\
\hline
		\multicolumn{2}{|l|}{Single-solution based metaheuristics:} & \\
    & Tabu search & \citep{ISI:000333870600009,Jin2012b,ISI:000418207900040,ISI:000419195500010,ISI:000320479300012,ISI:000290460100007,ISI:000261941400016,czapinski2013effective,ISI:000351482400005,ISI:000301216600010,ISI:000413300200004,ISI:000261555000011,ISI:000287350600005,ISI:000331498900004,Bozejko2016,Jin2011,Maischberger2011,VanLuong2013,dai2006multilevel,melab2011towards} \\
 \cline{2-3}
	  & Simulated annealing & \citep{ISI:000379511800009,ISI:000333870600009,ISI:000356000200004,ISI:000366052500005,ISI:000353746800009,ISI:000326045700018,ISI:000374711300002,ISI:000375042300022,Bozejko2009,Bozejko2016,Lazarova2008}\\
 \cline{2-3}
	  & Variable neigborhood search & \citep{ISI:000271571000075,ISI:000356110400002,davidovic2012mpi,ISI:000411486900004,ISI:000392861300012,ISI:000323470500012,ISI:000375207600001,ISI:000401878000007,ISI:000418434100005,aydin2008sequential,polacek2008cooperative} \\
 \cline{2-3}
	  & (Greedy randomized) adaptive search & \citep{Caniou2012,Santos2010}\\
 \cline{2-3}
    & Other single solution heuristics & \citep{Hifi2014} \\
		\hline
		\multicolumn{2}{|l|}{Population-based metaheuristics:} & \\
    & Genetic algorithm & \citep{ISI:000403021200004,ISI:000385330900006,ISI:000317884900003,ISI:000259761000009,ISI:000278526700026,ISI:000301155300035,ISI:000358469500006,ISI:000307760400022,Homberger2008,ISI:000266928200016,ISI:000321889500001,ISI:000385241600006,ISI:000345097100005,ISI:000388617500004,ISI:000387234200020,ISI:000271409100010,ISI:000398718400001,ISI:000324107400005,ISI:000312839600013,Lancinskas2013,Lancinskas2012,Lazarova2008,Sanci2011,Umbarkar2014,Wang2012,Zhao2011,VALLADA2009365,ISI:000415593600024} \\
 \cline{2-3}
	  & Other evolutionary algorithms & \citep{ISI:000305863300002,ISI:000417629800070,ISI:000406933400054,ISI:000317884900003,ISI:000374897200009,ISI:000276030200017,ISI:000341469100011,Banos2014,Nebro2010,Nowotniak2011,Redondo2008,Weber2011,Zhao2011,izzo2009parallel} \\
 \cline{2-3}
	  & Scatter search \& path relinking & \citep{ISI:000401889300020,bozejko2009solving}\\
 \cline{2-3}
	  & Ant colony optimization & \citep{ISI:000304221600003,ISI:000311921300005,ISI:000311921300006,ZHOU2017,hadian2012fine,ISI:000382768900001,cecilia2011parallelization,ISI:000384633000005,Abouelfarag2015,Lazarova2008,You2009,Zhao2011,ISI:000285048700020,Diego2012,Tsutsui2008,Dongdong2010}\\
 \cline{2-3}
    & Particle swarm optimization & \citep{aitzai2013parallel,ISI:000407732600020,ISI:000312839600013,ISI:000300080700009,rao2017solving,ISI:000407295800001,ISI:000302315500003,ISI:000272541300004,ISI:000293548900017,Deep2010,Ding2013,Wang2008}\\
 \cline{2-3}
    & Bee colony optimization & \citep{luo2014parallel,davidovic2011mpi,subotic2011different} \\
\cline{2-3}		
    & Fireworks algorithm & \citep{Ding2013} \\
		\hline
		\multicolumn{2}{|l|}{Hybrid metaheuristics} & \citep{ISI:000379511800009,ISI:000311921300006,ISI:000344552400009,ISI:000370099700013,ISI:000298631400006,ISI:000296012700016,ISI:000295018500007,ISI:000285413400014,ISI:000405549600001,ISI:000296539700061,ISI:000271404200004,ISI:000308548500016,ISI:000264988500031,Subramanian2010,ISI:000300080700009,ISI:000284747000034,ISI:000314737600028,ISI:000281591300145,Fujimoto2011,Ibri2010,Lancinskas2013,Taillard2012,Xhafa2008,Zhao2011,Zhu2009}\\
\hline		
		\multicolumn{2}{|l|}{Other heuristics} &  \citep{ISI:000349884400003,Sathe2012,ISI:000321869500004,Sanci2011}\\
\hline
		\multicolumn{2}{|l|}{Matheuristics} &  \citep{ISI:000355262900003,ISI:000290248600012}\\
		\hline
		\multicolumn{2}{|l|}{Multi-search algorithms} &  \citep{ISI:000299473600001,ISI:000402046200007,lahrichi2015integrative}\\
		\hline
	 \end{tabular}
	\end{addmargin}

\caption{\normalsize Parallel computational optimization studies in OR}
 \label{tab:parallelComputationalStudies}%
\end{table}%

In the remainder of this section, we present parallel computational optimization studies in OR grouped by algorithm types. An overview over the studies of our sample is given is Table \ref{tab:parallelComputationalStudies}.

{\bf Exact algorithms:} The majority of studies that apply exact algorithms parallelize branch-and-X algorithms. These studies analyze a broad range of optimization problems. Almost all adopt domain decomposition as parallelization strategy using a 1C/C or pC/C scheme with MPSS search differentiation, and most studies which report on the used communication topology apply a (one- or multiple-tier) master-slave approach. These efforts are not surprising as they reflect a straightforward (and traditional) way to parallelize branch-and-X algorithms. In contrast, the landscape of applied parallel programming models is more diverse and includes approaches based on threads, message passing, shared memory and GPGPUs. With regard to the former three models, mostly sublinear or linear speedup has been reported but there are also a few studies \citep{ISI:000404015200003,Borisenko2011,Galea2011} that report superlinear speedup. This speedup can be achieved, for example, when a parallel executed algorithm provides ``good" bounds that allow pruning large parts of the search tree at early stages. The use of GPGPUs has shown mixed results in terms of speedup; however, in some cases the reported speedup is substantial (between 76.96 and 170.69) \citep{ISI:000326057700005}, which makes GPGPUs highly appealing for parallelizing branch-and-X algorithms. However, it should also be acknowledged that several of these GPGPU studies have reported a high variance of speedup with regard to problem instances solved.         
Dynamic programming\footnote{An introduction into parallel dynamic programming is provided by \citet{almeida2006}.} is the second most often parallelized exact algorithm. Its parallelization in terms of addressed problems is quite diverse. In most cases, low-level is used as parallelization strategy with a 1C/RS scheme and SPSS search differentiation. The landscape of applied communication topologies is quite homogeneous, with almost all studies that report on the applied communication topology drawing on a (one- or multiple-tier) master-slave approach. In contrast, the set of implemented programming models is heterogeneous. Interestingly and in contrast to branch-and-X parallelization, the reported speedups are all sublinear. Studies that use GPGPUs report different ranges of speedup, with one study \citep{Tran2010} reporting an exceptionally high speedup in the range of 900-2,500.   
In addition, we found only a few studies which parallelize the interior point method. All of these studies address stochastic optimization problems, using low-level parallelism in a 1C/RS scheme with SPSS search differentiation, and they achieve sublinear or linear speedup. While all studies apply message passing as parallel programming model, the topologies used differ. Finally, a few exact methods designed for specific optimization problems (the knapsack problem \citep{ISI:000352249500002}, mixed integer linear programming \citep{Rossbory2013} and graph theory problems \citep{Kollias20142400,Bozdag2008}) have been parallelized. While all four studies show sublinear or linear speedup, the characteristics of algorithmic and computational parallelization are different.  

{\bf Single-solution based metaheuristics:}
Single-solution based metaheuristics manipulate and transform a single solution during the search. They can occur in many different forms and their parallelization has been discussed in \citep{Melab2006,talbi2009metaheuristics}. Parallelization can occur at the solution level, iteration level and algorithmic level. While parallelizing at the solution and iteration level generally corresponds to low-level parallelization with a 1C/RS scheme and SPSS search differentiation, parallelization at the algorithmic level is open to the broad range of parallelization strategies, and process and search control options.
Our literature review revealed that mainly three single-solution based metaheuristics have been parallelized: TS, SA and VNS.
TS has been applied to a variety of optimization problems. Most studies apply parallelization at the solution or iteration level, thereby adopting low-level parallelization with a 1C/RS scheme and SPSS search differentiation and a master-slave communication topology. We found a few exceptions from this algorithmic parallelization pattern; for example, \citet{Jin2012b,ISI:000261941400016,ISI:000331498900004,Jin2011} adopt cooperative multi-search parallelization of TS, and \citet{dai2006multilevel} implement domain decomposition parallelization of TS.
The landscape of applied parallel programming models is quite diverse and includes approaches based on threads, message passing, shared memory, SIMD, and GPGPUs.
 Speedup results are mixed, including superlinear speedup \citep{ISI:000320479300012,ISI:000287350600005}. The implementation on GPGPUs has shown substantial differences with regard to speedup, reaching values up to 420 \citep{czapinski2013effective}.
The landscape of parallel SA studies, which have also been applied to a variety of optimization problems, is more diverse than that of GA studies. It has been addressed by all four parallelization strategies with varying types of process and search control and with different programming models. In contrast to this heterogeneity, most studies apply a master-slave communication topology. Only a few studies report the achieved speedup, which is mostly sublinear. We found one study \citep{ISI:000326045700018} that parallelizes SA using GPGPU and achieves speedups in the range of about 73.44-269.46.
VNS has also been applied to many different problems with all four parallelization strategies and a variety of process and search control variations, communication topologies, and programming models. As in the case of SA, about half of the studies do not report on speedup and those which do report sublinear speedup, with the exception of \citet{polacek2008cooperative}, who achieve linear speedup. One study uses GPGPU \citep{ISI:000375207600001} and achieves a speedup in the range of 0.93-14.49.
Additionally, we found two studies \citep{Caniou2012,Santos2010} that parallelize (greedy randomized) adaptive search and one study \citep{Hifi2014} that parallelizes large neighborhood search (subsumed under ``other single solution heuristic (OSSH)" in Table \ref{tab:problem_algorithm}).

{\bf Population-based metaheuristics:}
In contrast to single-solution based metaheuristics, in population-based algorithms a whole population of solutions is evolved. Most prominent classes of population-based metaheuristics include evolutionary algorithms, scatter search and path relinking, swarm intelligence algorithms, and bee colony optimization \citep{talbi2009metaheuristics}. 
When population-based algorithms are parallelized, we distinguish three models which, albeit having been suggested originally for evolutionary algorithms in general and GAs in particular \citep{ISI:000178715400003,talbi2009metaheuristics,ISI:000222276100008, Melab2006,Cantu-Paz2005, Luque2005}, can be applied to other classes of population-based algorithms as well: \emph{global}, \emph{island (with or without migration)}, and \emph{cellular model}. In the global model, 
parallel techniques are used to speed up the operation of the algorithm without changing the basic operation of the sequential version. When the evaluation of the whole population is done in parallel, parallelism occurs at the iteration level; when the algorithm evaluates a single individuum in parallel, parallelism occurs at the solution level. In both cases, low-level parallelization applies. 
Island models typically run (identical or different) serial population-based algorithm on subpopulations to avoid getting stuck in local optima of the search space. If individuals can be transmitted between subpopulations, the island model is also referred to as \emph{migration model}; however, island models can also occur without migration. While in the former case, migration usually leads to a cooperative multi-search, the latter case generally corresponds to independent multi-search parallelization. The cellular model may be seen as a special case of the island model where an island is composed of a single individual. It should be noted that the models may be applied jointly (\citet{Cantu-Paz2005}, for example, describes such model combinations for GAs).

Evolutionary algorithms belong to the types of algorithms that have attracted substantial parallelization efforts. A good overview of the diversity with which combinations of different parallelization strategies and programming models can be applied to evolutionary algorithms is provided by \citet{ISI:000398718400001}. In our sample, we found a focus on GAs as a particular subclass of evolutionary algorithm; we subsume all evolutionary algorithms other than GAs under the residual subclass``other evolutionary algorithms".  
GAs have been parallelized for a variety of optimization problems. Most of the studies adopt the island model with migration (cooperative multi-search) with a pC/RS scheme and MPSS or MPDS search differentiation. Only a few studies use the island model without migration (independent multi-search) with a pC/RS scheme and MPSS search differentiation, or the global model (low-level) with a 1C/RS scheme and SPSS search differentiation. Interestingly, all but one study \citep{VALLADA2009365} apply synchronous communication. In the presence of the island model, a diversity of communication topologies has been applied with mostly message passing being used as programming model. In contrast, when the global model is applied, threads or GPGPU are drawn upon and mostly the master-slave topology is implemented. The described correlation between the parallelization strategy and the parallel programming model is not surprising as the communication between (a usually moderate number of) islands through exchanging messages is appealing while the processing of (a usually large number of) individuals in a global population through (an often large number of) threads executed on a CPU or GPGPU seems appropriate. Only about half of the 27 GA studies that we found report speedup values. Speedup results are overall mixed, including superlinear speedup \citep{Homberger2008,ISI:000388617500004}. The application of GPGPUs has led to homogeneous results, with a maximum speedup of about 33 \citep{Wang2012}.
Evolutionary algorithms other than GAs, such as differential evolution or immune algorithm, have been applied to a variety of optimization problems. Almost all of these studies adopt the island model with migration (cooperative multi-search) with a pC/RS scheme and MPSS or MPDS search differentiation. We found only two studies \citep{Banos2014,izzo2009parallel} which report an asynchronous communication. We identified no pattern regarding the applied communication topology and programming model. 

Swarm intelligence algorithms are inspired from the collective behavior of species such as ants, fish and birds. Subclasses of swarm intelligence algorithms for which we found parallelization studies are ant colony optimization (including ant colony systems and ``MAX-MIN Ant Systems" \citep{Dorigo2004}), particle swarm optimization, and fireworks algorithms. 
Parallelization strategies of ant colony optimization can be classified according to the above mentioned three strategies of parallelizing population-based metaheuristics; i.e., global, island or cellular model. Here, we follow the suggestion of \citet{randall2002parallel} to distinguish the parallel evaluation of solution elements, parallel ant colonies (independent or interacting) and parallel ants. These strategies are specializations of the global model, island model (without or with migration), and cellular model, respectively, of population-based metaheuristics. Interestingly, most of the parallelization studies using ant colony optimization have addressed the TSP. VRPs \citep{ISI:000285048700020,Diego2012} and assignment problems \citep{Tsutsui2008,Dongdong2010} have been solved by two studies each. Almost all studies use parallel ants or multiple ant colonies but, overall, the studies vary regarding parallelization strategies, process and search control, communication topologies and programming models. Those studies which qualify the achieved speedups, report sublinear speedups. The speedup achieved through GPGPU parallelization goes up to $25$. 
Particle swarm optimization has been applied to solve a diverse set of optimization problems. Most of the parallelization studies make use of the global or island model, realized as low-level or cooperative multi-search parallelization, respectively, with a master-slave communication topology. The process and search control implementations differ, with only one study \citep{Wang2008} reporting asynchronous communication. Mostly message passing and GPGPU are used as parallel programming model. Speedups achieved on GPGPU go up to about 190; studies not using the GPGPU model either do not report speedup values or show an overall diverse picture.  
In addition, we identified one study \citep{Ding2013} that applies a fireworks algorithm. 

{\bf Other population-based metaheuristics:}
We identified five studies that parallelize population-based metaheuristics other than evolutionary algorithms and swarm intelligence algorithms, namely scatter search and path relinking \citep{ISI:000401889300020,bozejko2009solving}, and bee colony optimization \citep{luo2014parallel,davidovic2011mpi,subotic2011different}. Addressed problems, algorithmic and computational parallelization characteristics as well as efficiency results (where reported) are quite diverse.

{\bf Hybrid metaheuristics:}
Hybrid metaheuristics are joint applications of several (meta)heuristics \citep{talbi2009metaheuristics,crainic2019parallel}. They are ``appropriate candidates" for the application of a(n) (independent or cooperative) multi-search strategy.
A diverse set of optimization problems has been investigated with parallel hybrid metaheuristics. The combinations of (meta)heuristics include ant colony optimization and local search, GAs and local search, GAs and SA, and GAs and TS, among others. Due to the diverse set of combined (meta)heuristics, unsurprisingly, the studies differ substantially with regard to addressed problems, parallelization strategies, process and search and control, communication topologies and parallel programming models. Although none of these studies report a superlinear speedup, \citet{Zhu2009} reports an achieved speedup of 403.91 when parallelizing a combination of ant colony optimization and pattern search with a GPGPU-based implementation. 

{\bf Problem-specific heuristics, other heuristics, matheuristics, and multi-search algorithms:}
Problem-specific heuristics have been parallelized for a variety of optimization problems, including a graph theory problem \citep{Dobrian2011}, TSPs \citep{ISI:000401878000008,Ismail2011}, a FSSP \citep{bozejko2009solving}, a facility location problem \citep{ISI:000361986700006}, a mixed integer linear program \citep{ISI:000418216800023}, and several other problems \citep{ISI:000386524100014,ISI:000349592500005,ISI:000349884400003,ISI:000255671800016,ISI:000378100800023,Luo2015}.
We found four studies which parallelize heuristics that differ from all types described above: an agent-based heuristic \citep{ISI:000349884400003}, an auction-based heuristic \citep{Sathe2012}, a Monte Carlo simulation inside a heuristic-randomization process \citep{ISI:000321869500004}, and a random search algorithm \citep{Sanci2011}. We found two studies which parallelize matheuristics \citep{ISI:000355262900003,ISI:000290248600012} and three studies which suggest multi-search algorithms \citep{ISI:000299473600001,ISI:000402046200007,lahrichi2015integrative}. Due to the diverse nature of the aforementioned studies, we do not look for patterns in algorithmic parallelization, computational parallelization and scalability results.  

\section{Research directions}
\label{sec:researchDirections}

 Based on the analysis of the identified literature published in the covered period (2008-2017), we subsequently suggest some research directions which may help (re)focusing on those areas that did not get much attention or were even neglected during the focused period. We would like to note that the observation of the absence or rareness of certain types of studies primarily refers to the aforementioned period. Work published prior to this period and surveys published earlier than this review (see Section \ref{sec:introduction}) have addressed some of the ``white spots" in research identified for the aforementioned period, which calls for \emph{re}-focusing on related research paths.

\subsection{Publication landscape and overall prospective research}

The analysis of publication data reveals that computational and parallel optimization in OR has been steadily attractive for many journals and conferences not only in the OR field but also in various neighbor disciplines. This broad interest is also reflected in the diverse landscape of which optimization problems have been solved by which (parallelized) algorithms. While this diversity shows the large relevance and broad applicability of computational parallelization in optimization, a closer look also reveals that the landscape is still fragmented despite the algorithmic accumulation of branch-and-X, GAs and TS studies and the problem accumulation of FSSPs, TSPs and VRPs. This makes it difficult to analyze which combinations of problems and algorithms are promising for parallelization and how the algorithmic and computational parallelization should be designed. It should be noted that in the presence of a broad scope of problems and algorithms in parallel optimization, the number of approximately 200 studies published in ten years is  relatively  low. Future research and education  can benefit from fostering  (knowledge on how to conduct) computational studies in parallel optimization to overcome the limitations imposed by fragmentation (recommendations 1a and 1b in Table \ref{tab:recommendations}).

\subsection{Object of parallelization} 

From the algorithmic perspective, branch-and-X algorithms represent the largest cluster of computational parallelization studies. In a few studies, this parallelization has even led to superlinear speedup but in most cases ``only" (sub)linear speedups have been achieved. Future research should shed more light on how to achieve superlinear speedups (recommendation 2a). With regard to dynamic programming, which is the second most often analyzed type of exact algorithms, the (sublinear) speedup achievements are less promising (see recommendation 2b). Again, our subsample of dynamic programming studies and their coding can serve as a basis for future investigations on more efficient dynamic programming parallelization, in particular on how to achieve superlinear speedup. We extend this recommendation to future research on parallelization of Lagrangean decomposition, which is -- as dynamic programming -- another methodology often used in the important field of stochastic optimization but which has hardly been parallelized. Parallelization efforts with regard to interior point methods are hardly existent, which asks for more research in this regard (recommendation 2c).

Among single-solution based metaheuristics, three metaheuristics have received particular attention regarding parallelization: TS, SA and VNS. For TS, speedup results are mixed, including two studies that report superlinear speedups, and the implementation on GPGPUs has shown substantial differences with regard to speedup. Future research should analyze this heterogeneous picture (recommendation 2d). With regard to SA and VNS, not much can be said on efficiency as, unfortunately, many studies do not report achieved speedups (see recommendation 2e). Beyond the aforementioned metaheuristics, other single-solution based metaheuristics, including \emph{greedy randomized adaptive search}, \emph{guided local search}, \emph{fast local search}, and \emph{iterated local search} \citep{gendreau2010handbook,gendreau2019handbook}, have not received much attention with regard to parallelization, which points to further research opportunities (recommendation 2f).

With regard to population-based metaheuristics, GAs are the most often parallelized type of algorithm. 
However, only a few studies provide speedup values, some of them reporting superlinear speedups. While these achievements are promising, not much  knowledge about the factors that lead to superlinear speedup (see recommendation 2g) has been developed. Furthermore, parallelization results for GAs as well as other evolutionary algorithms are mainly based on synchronous communication so that not much is known about the potential of applying asynchronous communication (recommendation 2h) . The second and third most often parallelized type of population-based metaheuristics are ant colony optimization and particle swarm optimization, respectively. With regard to ant colony optimization, achieved speedups are not very promising and mostly limited to applications to the TSP. Regarding particle swarm optimization, speedup results are quite mixed, with a promising speedup value of about 190 reported when using the GPGPU model. These results show that further research on parallelizing ant colony optimization and particle swarm optimization is  recommendable  (recommendation 2i).  Analogously to single-solution based metaheuristics, some algorithms of population-based methaheuristics, including SSPR, BCO and FA, have not received much attention, which shows avenues for further research (recommendation 2j).

Interestingly, we found only very few research on the parallelization of mat-heuristics. We believe that the parallelization of both of its' elements, metaheuristic components and exact mathematical programming techniques, are promising areas of future research (recommendation 2k).   

Similarly few attention has been attracted by multi-search algorithms, which offer a straightforward parallelization approach through parallelizing the execution of independent search algorithms involved in multi-search. We consider this research stream, in particular cooperative multi-search algorithms, to be highly relevant for future research on parallelization (recommendation 2l).
 
Beyond the previously identified algorithmic research directions, future research should also adopt problem-specific perspectives (recommendation 2m).

\subsection{Algorithmic parallelization and computational parallelization}

The algorithmic parallelization in the studies of our sample has drawn on all four (pure) parallelization strategies and on combinations of pure strategies. Low-level parallelization is the most often implemented strategy, with 83 out of 206 studies having used this type of parallelism. The process and search control is usually a 1C/RS scheme with SPSS search differentiation. Most studies which use low-level parallelism apply a master-slave communication topology, which is a straightforward approach. However, there are several exceptions, including fully-connected meshs (e.g., \citep{ISI:000417337900003}) and trees (e.g., \citep{ISI:000261892000011}). It would be useful to know under which conditions communication topologies other than the master-slave topology are advantageous for low-level parallelization (recommendation 3a). Interestingly, even for low-level parallelism a diverse set of parallel programming models and environments have been used, including message passing. This is a bit surprising as message passing is generally applied for the communication between "`heavy weight processes" executed on different computing nodes.

Domain decomposition as parallelization strategy occurs in 56 studies, with most of them parallelizing branch-and-X algorithms, which can be parallelized straightforward by decomposition. Regarding control cardinality, we found 1C and pC control modes applied similarly often. However, control and communication mostly follows an asynchronous, collegial scheme with no knowledge being exchanged between parallel processes; the used search differentiation is largely MPSS. Future research may explore opportunities that knowledge-based communication offer (recommendation 3b).

Independent multi-search as a parallelization strategy has been applied in only 18 studies, in contrast to cooperative multi-search, which has been implemented in 72 studies. This trend is encouraging as the potential of exchanging information between parallel processes in order to jointly achieve better solutions in less time has thereby been acknowledged by researchers. The vast majority of all studies which apply (independent or cooperative) multi-search uses a (synchronous) rigid synchronization (type ``RS"); we identified only four studies \citep{ISI:000290248600012,ISI:000351482400005,ISI:000331498900004,lahrichi2015integrative} which make use of knowledge-based communication. Future research should foster the exploration of knowledge-based communication when multi-search is applied (recommendation 3c).
-
Parallelization strategies can be combined to exploit complementary ways of parallelizations. For example, low-level and domain decomposition parallelism have been jointly applied to branch-and-X algorithms \citep{ISI:000368652500007,Adel2016} and to dynamic programming \citep{ISI:000385623100022}, and low-level and multi-search parallelism to genetic algorithms \citep{ISI:000324107400005,ISI:000271404200004}.
In total, we found eight  studies which apply such combinations. Future research should more intensively tap the potential that joint applications of different parallelization strategies offer (recommendation 3d). Finally, different parallelization strategies can be applied (separately) to the same algorithm and problem in order to compare their effectiveness and scalability and to determine most appropriate and inappropriate parallelizations.  Although we identified as many as 21 studies which follow this path, we encourage scholars to intensify  research in this regard (recommendation 3e).       

A broad range of different communication topologies has been applied, with master-slave being the most often used topology. The appropriateness of a communication topology needs to be linked to the particular algorithm and the applied parallelization strategy so that no general recommendations are appropriate. However, in the sample of computational studies we found only a few studies (e.g., \citep{mezmaz2014multi,Herrera2013,Rashid2010,aydin2008sequential}) that have implemented more than one topology for one parallelization strategy of a particular algorithm. This low number calls for more studies that investigate multiple topologies for particular combinations of algorithms and parallelization strategies (recommendation 3f). 

The parallel implementation of optimization algorithms has exploited overall a rich set of programming models and modern programming environments, including low-level threads (Java threads and POSIX threads), shared memory (mainly OpenMP), message passing (mainly MPI), and GPGPUs (mainly CUDA-based). In addition, also hybrid programming models, including message passing and shared memory, shared memory and GPGPU, threads and GPGPU, and message passing and threads, have been used in a few studies. Other programming models, such as SIMD, have only rarely been used. We found several studies which provide either no or incomplete information on the used parallel programming model(s). We recommend that studies report on the programming model and programming environment used for their parallelization (recommendation 3g). 

Only a few studies report on their (re-)use of software frameworks for parallelization, such as ParadisEO \citep{INRIA-dolphin} for parallel and distributed metaheuristics or Bob++ \citep{djerrah2006bob} for branch and bound parallelization. Reasons for not drawing on such frameworks can be manifold. Scholars may deliberately decide to not make use of them due to the inappropriateness of frameworks for their implementation case or due to too time-consuming efforts to get acquainted with the frameworks. Or, scholars are not aware of the existence of such frameworks. Either way, the development, propagation and use of re-usable software frameworks can substantially reduce the tedious and error-prone implementation of parallel optimization code (see recommendation 3h).        

\subsection{Performance of parallelization}

Scalability is essential regarding the appropriateness of a parallel implementation of an optimization algorithm. Interestingly, in 70 out of 206 studies speedup values are not (completely) reported or speedup is interpreted different from how it is usually done (see Section \ref{sec:parallel_performance_metrics}); for example, some studies determine the speedup by executing the serial and the parallel code on different hardware, resulting in speedup values  that are challenging to interpret. Other studies determine the speedup only of parts of an algorithm or use another parallel implementation as base (see \ref{sec:coding_parall_studies} for more details). In such cases, speedup values are hardly comparable with those of other studies and, thus, limit the usefulness of scalability analysis (see recommendation 4a). 

But even in case speedup is provided, comparisons with other studies need to be done carefully for several reasons: First, scalability results are difficult to compare with those of other studies when technological characteristics of parallel working units (or even of hardware environments) differ. For example, threads at the software level need to be distinguished from threads at the hardware level (hyperthreading), and MPI processes executed on different physical nodes may perform different from those executed on different cores on the same physical node. Second, values of weak speedup need to be distinguished from those of relative speedup (see Section \ref{sec:parallel_performance_metrics}). A list of issues related to speedup comparison is provided in \ref{sec:coding_parall_studies}. We condense our suggestions in recommendation 4b. 

We analyzed the studies in our sample with regard to how many parallel working units (threads or processes) have been used, which we refer to as \emph{range of parallelization}. The number of parallel threads executed on a CPU has been mostly not above 32 and it reaches its maximum at 128. When message passing is used on one or several nodes, the number of parallel processes units has in most cases not exceeded 256 and it has reached its maximum at 8,192. Hybrid approaches mostly use up to 1,024 parallel units, with the maximum number having been 2,048. Overall, the range of parallelization is quite limited compared to the number of parallel units that are available in modern parallel computing environments (see recommendation 4c). 

Our analysis of how studies in the literature have considered the effectiveness of parallelization (to obtain better solutions) showed that many studies do not analyze this category of performance and that those studies which provide effectiveness results use many different ways to report these. They apply different stop criteria (numbers of iterations, wall time, number of function evaluations, combinations of these criteria, etc.) and different evaluation criteria (objective value, relative gap to the best (known) solution value, numbers of instances solved to optimality, relative improvements, etc.), and often do not make the applied stop criteria explicit, which makes it difficult to assess parallel implementations and to compare studies with regard to effectiveness (see recommendation 4d). 

\subsection{Presentation of studies}

Finally, having reviewed more than two hundreds of parallelization studies, we found that studies differ substantially in the way how information on parallelization is provided, to what extent information is made explicit, and in which section(s) of the paper which information on parallelization is provided. This heterogeneity  may reflect different practices in various subfields and journals, and it not advisable to recommend any standardization in this regard. However, in several studies we found information on parallelization being reported incomplete, intransparent or distributed, which can make it tedious to fully understand the applied parallelization. The framework suggested in this paper may help to mitigate these issues when researchers adopt it and describe how it applies to their studies (recommendation 5).    

\begin{table}
\footnotesize
\begin{addmargin}{-3cm}
\centering
\begin{tabular}{|p{0.4cm}|p{17cm}|}
	\hline
	\multicolumn{2}{|c|}{\bf Publication landscape and overall prospective research}\\
	\hline
	1a &	Implementation of dedicated (tracks at) workshops and conferences and publication of edited books, such as \citep{Alba2005,talbi2006parallel}, and of special issues in journals\\ \hline
	1b & Integration of parallel optimization and its application in modern parallel computing environments in curricula of OR education\\	\hline
	\multicolumn{2}{|c|}{\bf Object of parallelization}\\
	\hline
	2a & Identification of those (algorithmic and computational) factors that drive superlinear speedup when parallelizing branch-and-X algorithms. The sample of 41 cases and their coding provided in this review offer a basis for this research.\\ \hline
	2b & Identification of ways to make parallelization of dynamic programming and of Lagrangean decomposition more efficient and to achieve superlinear speedup. Our subsample of dynamic programming studies and their coding can serve as a basis for future investigations.\\ \hline
	2c & Amplification of parallelization efforts with regard to interior point methods.\\ \hline
	2d & Analysis of heterogeneous picture of efficiency of TS parallelization to identify those factors that are most promising.\\ \hline
	2e & Amplification of scalability analysis with regard to parallelizations of SA and VNS.\\ \hline
	 2f &  Extension of parallelization efforts to a more comprehensive set of single-solution based metaheuristics, including \emph{greedy randomized adaptive search}, \emph{guided local search}, \emph{fast local search}, and \emph{iterated local search}.\\ \hline
	2g & Identification of those factors that drive superlinear speedup when parallelizing GAs.\\ \hline
	 2h & Application of asynchronous communication to genetic algorithms and other evolutionary algorithms.\\ \hline 	
	2i & Amplification of parallelization efforts with regard to ant colony optimization and particle swarm optimization.\\ \hline
	 2j &  Extension of parallelization efforts to a more comprehensive set of population-based metaheuristics, including \emph{scatter search \& path relinking}, \emph{bee colony optimization}, and \emph{fireworks algorithms}. \\ \hline
   2k &  Intensification of research on parallelizing matheuristics.\\ \hline
  2l & Intensification of research on the parallelization of multi-search algorithms, in particular those which include collaboration.\\ \hline
	
	2m & Adoption of problem-specific perspectives by analyzing which parallelization efforts (algorithms, parallel algorithm designs, parallel implementations) lead to which performance for a particular optimization problem. From Table \ref{tab:problem_algorithm} it can be seen that, in particular, FSSPs, TSPs, and VRPs  have attracted fairly high number of parallelization studies that can be used for further analysis.\\ \hline
	\multicolumn{2}{|c|}{\bf Algorithmic parallelization and computational parallelization}\\	\hline
  3a & Identification of conditions under which communication topologies other than the master-slave topology are advantageous for low-level parallelization.\\ \hline	
  3b & Exploration of opportunities that knowledge-based communication offers in the case of domain decomposition.\\ \hline	
  3c & Exploration of knowledge-based communication when multi-search parallelism is applied. \\ \hline	
  3d & Tapping the potential that joint applications of different parallelization strategies offer.\\ \hline	
  3e & Comparisons of effects that different parallelization strategies have when applied to a particular algorithm and problem in order to determine (in)appropriate parallelization strategies in this case.\\ \hline	
	3f &  Investigation of multiple strategies and/or multiple topologies for a particular algorithm in order to compare the performance of these alternatives.\\ \hline	
	3g & Documentation of programming model and programming environment used for parallelization.\\ \hline
	3h & Development and propagation of easy-to-use and flexible software frameworks for parallel optimization.\\ \hline
	\multicolumn{2}{|c|}{\bf Performance of parallelization}\\	\hline
	4a & Provision of values of both speedup and efficiency with regard to serial implementations executed on the same hardware.\\ \hline
	4b & Comparison of speedup and efficiency between algorithms of different studies needs to account for computational parallelization details and the type of speedup (e.g., relative or weak speedup) considered.
	\\ \hline
	4c & Extension of the range of parallelization (in terms of parallel computing units) to analyze scalability at larger levels.\\ \hline
	4d & Amplification of research on effectiveness of computational parallelization and documentation of applied stop and evaluation criteria.\\ \hline
	\multicolumn{2}{|c|}{\bf Presentation of studies}\\	\hline
  5 & Application of frameworks for describing parallelization studies to avoid incompleteness, intransparency and distributed provision of parallelization information. The framework suggested in this paper may be used.\\
	\hline
	 \end{tabular}
	\end{addmargin}
\caption{\normalsize Recommendations for future research on parallel computational optimization in OR}
 \label{tab:recommendations}%
\end{table}%

\section{Conclusion}
\label{sec:conclusion}

This invited review suggests a new integrative framework for parallel computational optimization. It integrates the perspectives on parallel optimization found in the disciplines of OR and computer science, and it distinguishes four levels: i) object of parallelization, ii) algorithmic parallelization, iii) computational parallization, and iv) performance of parallelization. We apply this framework to synthesize the body of literature (206 studies published between 2008 and 2017) of parallel computational optimization in OR. It should be noticed that the applicability of the suggested framework is not limited to the OR field.
Finally, we suggest several bundles of research recommendations for parallel computational optimization in OR, with the recommendations grouped along the layers of the suggested framework. 
 



\clearpage
\bibliography{x:/GUIDO/Literature/bibtex/authorsParallelComputing,x:/GUIDO/Literature/bibtex/bhpc,x:/GUIDO/Literature/bibtex/OR}

\normalsize

\appendix
\setstretch{1.5}

\section{Literature selection process}
\label{sec:app_literature_selection:process}

When invited by the editorial board of \emph{European Journal of Operational Research} in 2018, we were recommended to concentrate on the last decade of literature whenever possible. Following this recommendation is particularly reasonable for the body of literature on parallel optimization in OR because it accounts for a massive growth in computing performance in this period and resulting substantial advances of studies published regarding algorithmic parallelization, parallel software implementation and achieved computational results.

We conducted a title search in the most renowned OR journals. More specifically, we considered those 49 OR journals which are ranked ``A+", ``A", ``B" or ``C" in the German VHB-JOURQUAL 3 ranking of the German Academic Association for Business Research \citepappendix{VHB2015}; a complete list of these journals is included in Table \ref{tab:OR_journals}.
As we expected to find research related to parallel optimization in OR also in journals that are dedicated to parallel computing, we included the following four journals in our search: \emph{Journal of Parallel and Distributed Computing}, \emph{International Journal of Parallel Programming}, \emph{Parallel Programming} and \emph{Parallel Processing and Applied Mathematics}. We used \emph{Web of Science} to conduct a title search for both sets of journals, using the following search string:

\begin{quote}
(parallel* OR distributed OR "`shared memory" OR MPI OR OpenMP OR CUDA OR GPU OR SMP) AND NOT "parallel machine"      
\end{quote}
 
\begin{landscape}
\begin{table}%
\center
\begin{tabular}{ll}
	\hline
4OR	&	Journal of Decision Systems	\\
Annals of Operations Research	&	Journal of Economic Dynamics \& Control	\\
Artificial Intelligence	&	Journal of Forecasting	\\
Asia-Pacific Journal of Operational Research	&	Journal of Heuristics	\\
Central European Journal of Operations Research	&	Journal of Operations Management	\\
Computers and Operations Research	&	Journal of Revenue and Pricing Management	\\
Computers in Industry	&	Journal of Risk and Uncertainty	\\
Decision Sciences	&	Journal of Scheduling	\\
Decision Support Systems	&	Journal of the Operational Research Society	\\
Discrete Applied Mathematics	&	Logistics Research	\\
EURO Journal on Transportation and Logistics	&	Management Information Systems Quarterly	\\
European Journal of Operational Research	&	Managerial and Decision Economics	\\
Flexible Services and Manufacturing Journal	&	Manufacturing \& Service Operations Management	\\
Group Decision and Negotiation	&	Mathematical Methods of Operations Research	\\
IEEE Transactions on Systems, Man, and Cybernetics	&	Mathematical Programming	\\
IIE Transactions	&	Mathematics of Operations Research	\\
Information Systems Research	&	Naval Research Logistics	\\
INFORMS Journal on Computing	&	Operations Research	\\
Interfaces	&	Operations Research Letters	\\
International Journal of Forecasting	&	OR Spectrum	\\
International Journal of Information Technology \& Decision Making	&	SIAM Journal on Computing	\\
International Journal of Operations \& Production Management	&	System Dynamics Review	\\
International Journal of Operations Research	&	Transportation Research Part B: Methodological	\\
International Journal of Production Economics	&	Transportation Science	\\
International Journal of Production Research	&		\\
	\hline
\end{tabular}
\caption{Operation research journals considered in literature selection process (in alphabetical order)}
\label{tab:OR_journals}
\end{table}
\end{landscape}

Acknowledging that research on parallel optimization relevant to the OR discipline is likely to be published also in journals of other disciplines and in conference proceedings and books, we also conducted a title search using \emph{Web of Science Core Collection} without any restrictions regarding the publication outlet. However, we needed to adjust the search string in order keep the resulting list of articles manageable. The search strings that we used is as follows: 

\begin{itemize}
	\item ``parallel* optimization" OR ``parallel* branch" OR ``parallel* discrete" OR ``parallel heuristic" OR ``parallel exact" OR ``parallel meta" OR ``parallel genetic" OR ``parallel tabu" OR ``parallel evolutionary" OR ``parallel* ant colony" OR ``parallel* simulated annealing" OR ``parallel* variable neighborhood search" OR ``parallel* Greedy Randomized Adaptive Search Procedures" OR ``parallel* scatter search" OR ``parallel* dynamic programming"

	\item (MPI OR OpenMP OR CUDA OR GPU) AND (heuristic* OR exact OR meta OR genetic OR branch OR optimization OR discrete OR tabu)
	
	\item (parallel* AND algorithm) AND (knapsack OR transport OR logistics OR evolutionary)

\end{itemize}             

We also conducted a backward search of reference sections of literature reviews we identified (see the introduction of this article).    
 
Overall, our literature search returned more than 1,100 entries. With the support of a PhD and several student workers, we used the title of an article to decide whether it should be excluded from further analysis due to a missing fit with the scope of this review, resulting in a preliminary list of 238 entries . Finally, with the help of the student workers we analyzed the content of each of these articles and excluded further 83 entries for a variety of reasons, including a missing fit with scope and the use of languages other than English. Finally, we conducted a backward search of reference sections of the remaining 155 articles to mitigate the risk of overlooking relevant studies: in a first step, we selected potentially relevant articles based on their title; in a second step, we analyzed the selected articles by inspecting the full text to decide whether they should be included in the final set of considered articles or not; this procedure yielded 50 additional articles. Overall, the ultimate set of articles, referred to as \emph{our sample}, consists of 206 computational studies on parallel optimization in OR published between 2008 and 2017.  

\section{Coding of computational parallelization studies}
\label{sec:coding_parall_studies}

This section contains the detailed coding results of our sample with the exception of three studies: \citetappendix{Oestermark2014,Oestermark2015} do not explicit the algorithm parallelized; \citetappendix{ISI:000301216600031} parallelizes the problem-specific evaluation of objective function but no overall algorithm is considered. To sum up, the tables in this section include 203 studies of the full sample (206 studies).

The articles are grouped along types of algorithms, with Table \ref{tab:results_exact} addressing exact methods, Table \ref{tab:results_singleSolHeur} addressing single-solution based metaheuristics, Table \ref{tab:results_populationBasedHeur} addressing population-based metaheuristics, Table \ref{tab:results_hybridHeur} addressing hybrid metaheuristics, and Table \ref{tab:results_prblSpecificOtherHeurMH} addressing problem-specific heuristics, other heuristics, and matheuristics. Unsurprisingly, not all studies included in our sample provide sufficiently precise details that allow coding all attributes. In cases where incomplete or ambiguous information is provided
, we use the value ``n/a". We need to point to two exceptions from this rule: 1) in the column ``Process and search control", which show a triple classification, the usage of ``n/a" for one or more of the three classes may confuse the reader. Thus, we prefer to use the symbol ``?" where information is not available or ambiguous, or where our classification is not applicable (e.g., in reference
\citep{ISI:000341469100011}, a semi-synchronous mode is used because MPI-synchronization occurs at a pairwise level but not at a global level (p. 15)). 2) The entry ``n/a" in the ``Scalability" column has a more sophisticated interpretation, which we unfold in the text below. 

The entries in the columns labeled ``Problem" and ``Algorithm" use the abbreviations as shown in Table \ref{tab:problem_algorithm} in the main text of the article. Entries in columns labeled ``Parallelization strategy", ``Process \& search control", ``Communication topology" and ``Programming model" are used as described in the main text.
     
The column ``Scalability" covers both speedup and efficiency. It shows different types of entries: speedup that is qualified by its type of efficiency is provided in the form ``sublinear (n=2-16)", for example, where the range of n indicating the numbers of parallel processing units used. Speedup that varies between (sub)linear and superlinear depending on tested instances is described accordingly. Speedup achieved with GPGPUs is given as a single value or as an interval. We do not qualify speedup in this case as the number of parallel working units (usually GPGPU threads) needs to be interpreted different from that counting other parallel working units (CPU threads, processes) because they differ substantially from a technological perspective. Also, for the same reason, the determination of efficiency of parallelization should not be computed as the ratio of speedup and the number of parallel processing units. The entry ``n/a" in the ``Scalability" column is an umbrella type and can have several different meanings described below. When more than one experiment has been conducted (e.g, applying different (versions of) algorithms, different (sets of) benchmark instances, and/or different programming models), speedup information is numbered.    

Reasons for labeling scalability as ``n/a" turned out to be appropriate for manifold reasons:

\begin{itemize}
	\item Times are compared with theoretical serial times.
	\item Speedup is related to other parallel executed algorithms or to parallel execution of the same algorithm (for example, because the execution on a single processing unit was practically infeasible due to time limitations); i.e., we report only speedups (weak or relative) related to serial executions of algorithms.
	\item The type of reference execution is unknown.
	\item No speedup values are reported or tedious work is necessary to determine them from data reported.
	\item Speedup values are provided in in supplementary material which is inaccessible.
	\item Speedup values only refer to parts of algorithms.
	\item Running times must not be compared as i) different (hardware) machines/computing environments are used, or ii) different levels of objective functions are achieved by reference execution(s) and execution of parallel algorithm.
	\item Parallelization is conducted in a virtual environment where no physical parallelization occurs. Then,  execution times are hardly comparable as parallel execution times will often be larger than sequential times due to parallelization overhead.
\end{itemize} 
 
We do not qualify speedup (as ``linear", for example) in the case of GPGPU as programming model as the number of parallel working units (usually GPGPU threads) needs to be interpreted different from that counting other parallel working units (CPU threads, processes) because they differ substantially from a technological perspective. Also, for the same reason, the determination of efficiency of parallelization should not be computed as the ratio of speedup and the number of parallel processing units.

\tiny
\begin{landscape}
\begin{longtable}{|p{3.5cm}p{1.2cm}p{1.2cm}p{2.8cm}p{2.3cm}p{1.7cm}p{2.5cm}p{4.2cm}|}
\hline
\multirow{2}{*}{\textbf{Reference}} & \multirow{2}{*}{\textbf{Problem}} & \multirow{2}{*}{\textbf{Algorithm}} & {\textbf{Parallelization}} & {\textbf{Process \&}} & {\textbf{Communication}} & {\textbf{Programming }} & \multirow{2}{*}{\textbf{Scalability}}\\
& & & {\textbf{strategy}} & {\textbf{search control}} & {\textbf{topology}} & {\textbf{model}} &\\ 
\hline
\endhead
\hline
\endfoot
\hline\\
 \caption{\normalsize Computational parallelization studies (exact algorithms)}
\endlastfoot

\citep{mezmaz2014multi} & {FSSP} & {B-a-X} & domain decomposition &  1. 1C/C/MPSS, 2. pC/C/MPSS & {1.master-slave, 2. ring} & {n/a} & n/a \\
\citep{ISI:000318042500009} & {FSSP} & {B-a-X} & {low-level} & {1C/RS/SPSS} & {master-slave} & {GPGPU} & [40.52-60.64] \\
\citep{Herrera2017} & {BFP} & {B-a-X} & {domain decomposition} & pC/C/MPSS & {n/a} & {threads} & 1. sublinear (n=1-16), 2. and 3. linear (n=1-16) \\
\citep{ISI:000255647700028} & {GTP} & {B-a-X} & {domain decomposition}  & pC/C/MPSS & {mesh} & {message passing} & linear (n=4-16) \\
\citep{ISI:000404015200003} & {Other} & {B-a-X} & {domain decomposition} &  1C/C/MPSS & {master-slave} & {hybrid (message passing + shared memory)} & superlinear (n=160) \\
\hdashline
\citep{ISI:000332452400025} & KP    & {B-a-X} & domain decomposition & pC/C/MPSS & tree  & message passing & [1.9-7.3] (unreported n) \\
\citep{ISI:000299550000011} & Other & {B-a-X} & domain decomposition  & 1C/C/MPSS & n/a   & 1. shared memory, 2. message passing & sublinear (n=2-16) \\
\citep{ISI:000321416800013} & GTP   & {B-a-X} & domain decomposition & 1C/C/MPSS & n/a   & threads & sublinear (n=2-16) \\
\citep{ISI:000354949200009} & GTP   & {B-a-X} & domain decomposition  & 1C/C/MPSS & n/a   & threads & linear (n=4) \\
\citep{eckstein2015pebbl} & Other & {B-a-X} & domain decomposition & 1C/C/MPSS & master-slave & message passing & linear (n=2-8192) \\
\hdashline
\citep{ISI:000335544500016} & MILP  & {B-a-X} & 1. indep. multi-search, 2./3. coop. multi-search & 1. pC/RS/SPDS, 2. pC/KC/SPDS, 3. pC/?/SPDS & master-slave & message passing & n/a \\
\citep{ISI:000395014100005} & Other & {B-a-X} & domain decomposition & 1C/C/MPSS & master-slave & GPGPU & [1.67-5.79] \\
\citep{ISI:000398718400009} & FSSP, TSP, Other & {B-a-X} & domain decomposition & 1C/C/MPSS & n/a   & GPGPU & n/a \\
\citep{ISI:000322300300002} & Other & {B-a-X} & domain decomposition & 1C/C/MPSS & master-slave & shared memory (read/write lock used) & linear (n=5) \\
\citep{Bak2011a} & Other & {B-a-X} & domain decomposition & 1C/C/MPSS & master-slave & message passing & varies between instances (n=2-20) \\
\hdashline
\citep{ISI:000390638300008} & {FSSP} & {B-a-X} & domain decomposition & 1C/C/MPSS & n/a   & GPGPU & n/a \\
\citep{ISI:000351682800004} & Other & {B-a-X} & domain decomposition & pC/C/MPSS & tree  & hybrid (shared memory + message passing) & varies between instances (n=16), sublinear/linear (n=32-64) \\
\citep{barreto2010parallel} & MILP  & {B-a-X} & domain decomposition & 1C/C/MPSS & master-slave & 1. message passing, 2. shared-memory & sublinear (n=2-100) \\
\citep{ISI:000368652500007} & {FSSP} & {B-a-X} & hybrid (low-level + domain decomposition) & n/a & fully connected mesh & 1. message passing, 2. hybrid (shared memory + message passing) & sublinear/linear (n=1-16 GPUs, m=1-512 CPUs) \\
\citep{ISI:000342481400006} & {FSSP} & {B-a-X} & domain decomposition & pC/C/MPSS & n/a   & 1. n/a, 2. threads, 3. n/a, 4. n/a & 1. [80-160], 2.linear (n=2-6 CPUs), 3. [79.42-124.1], 4.[198.55-222.02] \\
\hdashline\citep{paulavivcius2009parallel} & Other & {B-a-X} & low-level  & 1C/RS/SPSS & master-slave & shared memory & sublinear (n=2-4) \\
\citep{ISI:000258760900007} & KP    & {B-a-X} & domain decomposition & pC/C/MPSS & tree  & message passing & sublinear (n=8-32) \\
\citep{ISI:000326057700005} & {FSSP} & {B-a-X} & domain decomposition & pC/C/MPSS & n/a   & GPGPU & [76.96-170.69] \\
\citep{aitzai2013parallel} & {JSSP} & {B-a-X} & domain decomposition & 1.pC/C/MPSS, 1C/C/MPSS & hybrid (master-slave + ring) & message passing  & varies between instances (n=3-4) \\
\citep{ISI:000401878000008} & TSP   & {B-a-X} & domain decomposition & 1C/C/MPSS & master-slave & message passing  & n/a \\
\citep{ISI:000289069700006} & Other & {B-a-X} & domain decomposition & pC/C/MPSS & hybrid (master-slave + ring) & message passing  & n/a \\
\citep{ISI:000268558800004} & MILP  & {B-a-X} & domain decomposition & 1C/C/MPSS & master-slave & message passing  & n/a \\
\citep{ISI:000392789400017} & SOP   & {B-a-X} & domain decomposition & pC/C/MPSS & tree  & hybrid (message passing + threads) & n/a \\
\hdashline
\citep{ISI:000352667900008} & MILP  & {B-a-X} & domain decomposition & 1C/C/MPSS & master-slave & message passing & sublinear (n=12) \\
\citep{ISI:000319238800010} & SOP   & {B-a-X} & domain decomposition & 1C/C/MPSS & n/a   & message passing & sublinear (n=8-32) \\
\citep{ISI:000417337900003} & SOP   & IPM   & low-level & 1C/RS/SPSS & fully connected mesh & message passing & sublinear (n=8-64) \\
\citep{Dias2013a} & Other & DP    & low-level & 1C/RS/SPSS & master-slave & message passing & sublinear (n=8-256) \\
\citep{ISI:000362899100010} & SOP   & DP    & 1. low-level, 2. coop. multi-search & 1. 1C/RS/SPSS, 2.pC/RS/SPDS & 1. tree, 2. master-slave & message passing & sublinear (n=12) \\
\hdashline
\citep{ISI:000385623100022} & Other & DP    & hybrid (low-level + domain decomposition)  & pC/RS/MPDS & n/a   & 1. shared memory, 2. distributed memory & sublinear (n=(1 or 8)-128) (probably wrt. distributed memory) \\
\citep{ISI:000261892000011} & n/a   & DP    & low-level & 1C/RS/SPSS & tree  & threads & sublinear (n=4-64) \\
\citep{ISI:000278924200005} & KP, GTP, Other & DP    & coop. multi-search & pC/C/SPDS & n/a   & shared memory & mostly sublinear (n<=32) \\
\citep{boyer2012solving} & KP    & DP    & low-level & 1C/RS/SPSS & tree  & GPGPU & [18.9-26.05] \\
\citep{ISI:000380751900011} & Other & DP    & low-level & 1C/RS/SPSS & {tree} & {hybrid (shared memory + GPGPU)} & n/a \\
\hdashline
\citep{Kollias20142400} & {GTP} & PSEA  & low-level  & 1C/RS/SPSS & master-slave & hybrid & linear (n<=1024) \\
\citep{Bozdag2008} & {GTP} & PSEA  & domain decomposition & pC/RS/MPSS & mesh  & message passing & sublinear (n<=40) \\
\citep{ISI:000352249500002} & KP    & PSEA  & low-level & n/a & n/a   & 1. shared memory, 2. GPGPU & 1. sublinear or linear (varies between instances) (n=2-16), 2. [2-16] \\
\citep{Adel2016} & JSSP   & {B-a-X} & hybrid (low-level+domain decomposition) & n/a & master-slave & GPGPU  & [0.31-65.5] \\
\citep{Borisenko2011} & Other & {B-a-X} & domain decomposition & 1C/C/MPSS & {n/a} & hybrid (shared memory + message passing) & superlinear (n<=32) \\
\hdashline
\citep{Boukedjar2012} & KP    & {B-a-X} & domain decomposition & 1C/C/MPSS & {n/a} & GPGPU & [3.84-9.27] \\
\citep{Carneiro2011} & TSP   & {B-a-X} & domain decomposition & pC/C/MPSS & {n/a} & 1. shared memory, 2. GPGPU & 1. sublinear (n=4), 2. [7.61-10.69] \\
\citep{Galea2011} & AP    & B-a-X & 1. low-level, 2. domain decomposition & 1. 1C/RS/SPSS, 2. n/a & ring  & 1. threads, 2. SIMD & superlinear (n<=24) \\
\citep{Herrera2013} & Other & {B-a-X} & domain decomposition & 1. 1C/C/MPSS, 2. pC/C/MPSS & 1. master-slave, 2. tree & hybrid (message passing + threads) & sublinear (n=32-128) \\
\citep{Hong2010} & {Other} & IPM   & low-level & 1C/RS/SPSS & n/a   & message passing & linear (n<=16) \\
\citep{Kumar2011} & GTP   & DP    & 1. n/a, 2. low-level  & 1. n/a, 2. 1C/RS/SPSS & {n/a} & {1. threads, 2. GPGPU} & 1. sublinear (n=2), 2. [10.27-13.82] \\
\citep{Lubin2012} & SOP   & IPM   & low-level  & 1C/RS/SPSS & n/a   & message passing & sublinear (n=32-2048) \\
\citep{Lucka2008} & Other   & IPM   & low-level & 1C/RS/SPSS & master-slave & message passing & sublinear (n<=16) \\
\hdashline
\citep{Rashid2010} & KP    & DP    & low-level  & 1C/RS/SPSS & 1. master-slave, 2. mesh & shared memory & sublinear (n<=16) \\
\citep{Rossbory2013} & MILP  & PSEA  & domain decomposition  & 1C/RS/SPSS & master-slave & n/a   & sublinear (n=12) \\
\citep{Estrada2011} & BFP, Other & B-a-X & domain decomposition  & pC/C/MPSS & n/a   & threads & linear (1<n<4), sublinear (4<n<128), n=average no. of running threads \\
\citep{Tran2010} & GTP   & DP    & low-level  & 1C/RS/SPSS & n/a   & GPGPU & two algorithm versions: 1. [48-52], 2. [900-2500]

 	\label{tab:results_exact}%
\end{longtable}%
\end{landscape}
\normalsize

\tiny
\begin{landscape}
\begin{longtable}{|p{3.5cm}p{1.2cm}p{1.2cm}p{2.8cm}p{2.3cm}p{1.7cm}p{2.5cm}p{4.2cm}|}
\hline
\multirow{2}{*}{\textbf{Reference}} & \multirow{2}{*}{\textbf{Problem}} & \multirow{2}{*}{\textbf{Algorithm}} & {\textbf{Parallelization}} & {\textbf{Process \&}} & {\textbf{Communication}} & {\textbf{Programming }} & \multirow{2}{*}{\textbf{Scalability}}\\
& & & {\textbf{strategy}} & {\textbf{search control}} & {\textbf{topology}} & {\textbf{model}} &\\ 
\hline
\endhead
\hline
\endfoot
\hline\\
 \caption{\normalsize Computational parallelization studies (single-solution based metaheuristics)}
\endlastfoot

\citep{ISI:000379511800009} & JSSP   & SA & 1a. low-level, 1b. coop. multi-search, 2. coop. multi-search  & 1a. 1C/RS/SPSS, 1b. pC/C/?PDS, 2. pC/C/SPDS & n/a   & shared memory & n/a \\
\citep{ISI:000333870600009} & MSP   & TS    & low-level & 1C/RS/SPSS & master-slave & threads & sublinear (n=8) \\
\citep{ISI:000333870600009} & MSP   & SA    & indep. multi-search & pC/RS/SPDS & master-slave & threads & sublinear (n=8) \\
\citep{ISI:000271571000075} & {JSSP} & {VNS} & low-level & 1C/RS/SPSS & {master-slave} & {n/a} & n/a \\
\citep{ISI:000356110400002} & {FSSP} & {VNS} & coop. multi-search & pC/C/MPDS & {bidirectional (two nodes)} & {message passing} & sublinear (n=2) \\
\hdashline
\citep{davidovic2012mpi} & {MSP} & {VNS} & 1. low-level, 2. domain decomposition, 3. coop. multi-search, 4. coop. multi-search; 5. coop. multi-search & 1. 1C/RS/SPSS, 2. 1C/C/MPSS, 3. pC/C/SPDS, 4. pC/C/SPDS, 5. pC/C/MPDS & {1.- 2. n/a, 3.-4. master-slave, 5.ring} & {message passing} & n/a \\
\citep{ISI:000411486900004} & {KP} & {VNS} & domain decomposition & 1C/RS/MPDS & {master-slave} & {message passing} & n/a \\
\citep{ISI:000392861300012} & {Other} & {VNS} & low-level & 1C/RS/SPSS & {master-slave} & {threads} & sublinear (n=8) \\
\citep{ISI:000323470500012} & {Other} & {VNS} & low-level & 1C/RS/SPSS & {master-slave} & {threads} & n/a \\
\citep{ISI:000375207600001} & {VRP} & {VNS} & low-level & 1C/RS/SPSS & {n/a } & {GPGPU} & [0.93-14.49] \\
\hdashline
\citep{ISI:000401878000007} & {VRP} & {VNS} & n/a & pC/C/MPSS & {star} & {n/a} & sublinear (n=6) \\
\citep{ISI:000418434100005} & {VRP} & {VNS} & low-level & 1C/RS/SPSS & {n/a} & {shared memory} & sublinear (n=2-32) \\
\citep{ISI:000356000200004} & FSSP   & SA    & domain decomposition & 1C/KS/SPMS & master-slave & message passing & n/a \\
\citep{ISI:000366052500005} & VRP   & SA    & coop. multi-search & pC/RS/SPDS & master-slave & threads & n/a \\
\citep{ISI:000353746800009} & VRP   & SA    & coop. multi-search & pC/RS/SPDS & master-slave & threads & n/a \\
\hdashline
\citep{ISI:000326045700018} & Other & SA    & 1. indep. multi-search, 2. coop. multi-search & pC/RS/SPDS & 1. unconnected, 2. master-slave & GPGPU & [73.44-269.46] \\
\citep{ISI:000374711300002} & BFP   & SA    & coop. multi-search & pC/?/?PDS & n/a   & message passing & sublinear (n=24-192) \\
\citep{ISI:000375042300022} & VRP   & SA    & 1. domain decomposition, 2. cooperative multi-search & 1. 1C/KS/SPDS, 2. pC/C/MPDS & master-slave & 1. shared memory, 2. message passing & n/a \\
\citep{Jin2012b} & VRP   & TS    & coop. multi-search & pC/RS/SPDS & master-slave & threads & n/a \\
\citep{ISI:000418207900040} & JSSP   & TS    & low-level & 1C/RS/SPSS & master-slave & shared memory & sublinear (n=2-224) \\
\hdashline
\citep{ISI:000419195500010} & Other & TS    & low-level & 1C/RS/SPSS & n/a   & GPGPU & [4] \\
\citep{ISI:000320479300012} & FSSP   & TS    & low-level & 1C/RS/SPSS & master-slave & SIMD & linear or superlinear (varies between instances) (n=2) \\
\citep{ISI:000290460100007} & FSSP   & TS    & low-level & 1C/RS/SPSS & master-slave & GPGPU & [<=89] \\
\citep{ISI:000261941400016} & AP    & TS    & coop. multi-search & pC/C/MPDS & communication via memory & shared memory & sublinear (n=10) \\
\citep{czapinski2013effective} & AP    & TS    & 1. low-level, 2. coop. multi-search & 1C/RS/SPSS & master-slave & GPGPU & [<=420] \\
\citep{ISI:000351482400005} & Other & {TS} & coop. multi-search & pC/KS/MPSS & {n/a} & 1. threads, 2. GPGPU & 1. linear (n=4), 2. [7.12-50.4] \\
\citep{ISI:000301216600010} & VRP   & {TS} & low-level & 1C/RS/SPSS & {n/a} & message passing & sublinear (n=10-80) \\
\citep{ISI:000413300200004} & FSSP   & {TS} & low-level & 1C/RS/SPSS & {n/a} & GPGPU & [1.15 -139.87] \\
\citep{ISI:000261555000011} & TSP,FSSP & {TS} & low-level & 1C/RS/SPSS & {master-slave} & GPGPU & sublinear (n=4) \\
\hdashline
\citep{ISI:000287350600005} & JSSP   & {TS} & indep. multi-search & pC/?/? & {n/a} & n/a   & two algorithm versions: 1. superlinear (n<2-20), 2.linear (n=2-20) \\
\citep{ISI:000331498900004} & VRP   & {TS} & coop. multi-search & pC/KC/MPDS & {master-slave} & message passing & n/a \\
\citep{Bozejko2009} & JSSP   & {SA} & low-level & 1C/RS/SPSS & master-slave & SIMD & [1.7-5.6] \\
\citep{Bozejko2016} & FSSP   & {TS} & low-level & 1C/RS/SPSS & {n/a} & message passing & sublinear (n=2-10) (no information on which algorithm was analyzed) \\
\citep{Bozejko2016} & FSSP   & {SA} & low-level & 1C/RS/SPSS & {n/a} & message passing & sublinear (n=2-10) (no information on which algorithm was analyzed) \\
\hdashline
\citep{Caniou2012} & Other & {GRAS} & indep. multi-search & pC/RS/MP?S & {n/a} & n/a   & sublinear/linear (n=32-256] \\
\citep{Hifi2014} & KP    & {OSSH} & domain decomposition & pC/KC/SPDS & {master-slave} & message passing & n/a \\
\citep{Jin2011} & VRP   & TS    & coop. multi-search & pC/C/MPDS & {master-slave} & message passing & n/a \\
\citep{Lazarova2008} & TSP   & SA    & 1. coop. multi-search, 2. ?, 3. coop. multi-search & 1. pC/?/MPDS, 2. ?, 3. pC/RS/MPDS & master-slave & hybrid (message passing + shared memory) & results vary between instances (n=2-10) \\
\citep{Maischberger2011} & VRP   & TS & low-level & 1C/RS/SPSS & {n/a} & n/a   & n/a \\
\hdashline
\citep{Santos2010} & Other & GRAS  & domain decomposition & 1C/RS/SPSS & master-slave & GPGPU & [1.14-13.89] \\
\citep{VanLuong2013} & 1. AP, 2. BFP, 3. TSP & TS    & low-level  & 1C/RS/SPSS & master-slave & GPGPU & 1. [0.5-18.6], 2. [39.2-243], 3. [0.6-19.9]\\
\citep{aydin2008sequential} & JSSP & VNS & coop. multi-search & 1. pC/RS/SPDS, 2.pC/C/MPDS, 3.pC/C/MPDS, 4.pC/C/MPDS & 1. master-slave, 2. master-slave, 3. ring, 4. mesh & message passing  & n/a\\
\citep{dai2006multilevel} & Other & TS & 1. low-level, 2. domain decomposition & 1. 1C/RS/SPSS, 2. pC/C/MPSS & 1. master-slave, 2. n/a & n/a   & 1. sublinear, 2. n/a\\
\citep{melab2011towards} & AP & TS & low-level & 1C/RS/SPSS & master-slave & GPGPU   & [0.9-15.1] \\
\hdashline
\citep{polacek2008cooperative} & VRP & VNS & coop. multi-search & pC/C/MSDS & master-slave & message passing   & linear (n=2,...,32)

 	\label{tab:results_singleSolHeur}%
\end{longtable}%
\end{landscape}
\normalsize

\tiny
\begin{landscape}
\begin{longtable}{|p{3.5cm}p{1.2cm}p{1.2cm}p{2.8cm}p{2.3cm}p{1.7cm}p{2.5cm}p{4.2cm}|}
\hline
\multirow{2}{*}{\textbf{Reference}} & \multirow{2}{*}{\textbf{Problem}} & \multirow{2}{*}{\textbf{Algorithm}} & {\textbf{Parallelization}} & {\textbf{Process \&}} & {\textbf{Communication}} & {\textbf{Programming }} & \multirow{2}{*}{\textbf{Scalability}}\\
& & & {\textbf{strategy}} & {\textbf{search control}} & {\textbf{topology}} & {\textbf{model}} &\\ 
\hline
\endhead
\hline
\endfoot
\hline\\
 \caption{\normalsize Computational parallelization studies (population-based metaheuristics)}
\endlastfoot

\citep{ISI:000285048700020} & VRP   & {ACO} & coop. multi-search & pC/RS/MP?S & ring  & n/a   & n/a \\
\citep{ISI:000304221600003} & TSP   & {ACO} & coop. multi-search & pC/C/MP?S & communication partners selected dynamically & message passing & sublinear (n=5-35) \\
\citep{ISI:000311921300005} & TSP   & {ACO} & low-level & 1C/RS/SPSS & mesh  & GPGPU & [~0.6-22] \\
\citep{ISI:000311921300006} & TSP   & {ACO} & low-level & 1C/RS/SPSS & master-slave & GPGPU & a. [2-19.47], b. [0.06-23.6] \\
\citep{ZHOU2017} & TSP   & {ACO} & low-level & 1C/RS/SPSS & n/a   & hybrid (shared memory (task-based) + SIMD) & linear (n=2-16) \\
\hdashline
\citep{hadian2012fine} & TSP   & {ACO} & low-level & 1C/RS/SPSS & master-slave & threads & sublinear (n=2-24) \\
\citep{ISI:000382768900001} & TSP   & {ACO} & coop. multi-search & pC/RS/MPDS & hybrid (master-slave + fully connected mesh) & message passing & sublinear (n=2-8) \\
\citep{cecilia2011parallelization} & TSP   & {ACO} & low-level & 1C/RS/SPSS & master-slave & GPGPU & n/a \\
\citep{ISI:000384633000005} & TSP   & {ACO} & low-level & 1C/RS/SPSS & {master-slave} & {GPGPU} & [5-25] \\
\citep{Abouelfarag2015} & TSP   & {ACO} & low-level & 1C/RS/SPSS & master-slave & shared memory & sublinear (n=2-32) \\
\hdashline
\citep{luo2014parallel} & {BFP} & {BCO} & coop. multi-search & pC/RS/MPDS & several fully connected meshs (isolated from each other) & {GPGPU} & [13-56] \\
\citep{aitzai2013parallel} & {JSSP} & {PSO} & domain decomposition & 1C/RS/MPSS & master-slave & message passing  & n/a \\
\citep{ISI:000403021200004} & Other & {GA} & low-level & 1C/RS/SPSS & master-slave & threads & n/a \\
\citep{ISI:000305863300002} & Other & {OEA} & low-level & 1C/RS/SPSS & master-slave & GPGPU & [1.3-14] \\
\citep{ISI:000385330900006} & Other & {GA} & coop. multi-search & pC/RS/MPDS & {ring} & message passing & n/a \\
\hdashline
\citep{ISI:000407732600020} & {Other} & {PSO} & hybrid (low-level + coop. multi-search) & pC/?/MPDS & {hybrid (master-slave + ring)} & {message passing} & [3-18] (n=5) \\
\citep{ISI:000417629800070} & Other & {OEA} & coop. multi-search & pC/RS/MPSS & master-slave & n/a   & superlinear (n<=14) \\
\citep{ISI:000406933400054} & Other & {OEA} & coop. multi-search & pC/RS/MPSS & grid  & message passing & sublinear (n=360) \\
\citep{ISI:000317884900003} & MSP   & {GA} & coop. multi-search & pC/RS/MPDS & fully connected mash & threads & varies between algorithms and instances (n=4-8) \\
\citep{ISI:000317884900003} & MSP   & {OEA} & coop. multi-search & pC/RS/MPDS & fully connected mash & threads & varies between algorithms and instances (n=4-8) \\
\hdashline
\citep{ISI:000374897200009} & Other & {OEA} & low-level & 1C/RS/SPSS & master-slave & threads & sublinear/linear (n<=24) \\
\citep{ISI:000276030200017} & Other & {OEA} & indep. multi-search & pC/RS/MPSS & {master-slave} & {message passing} & n/a \\
\citep{ISI:000341469100011} & Other & {OEA} & coop. multi-search & pC/?/DP?S & double-linked list & message passing & sublinear/linear (n<=128) \\
\citep{ISI:000259761000009} & Other & {GA} & coop. multi-search & pC/RS/MPDS & 1.-2. ring, 3.-4. mesh, 5. fully connected mesh, 6. randomly connected mesh & message passing & n/a \\
\citep{ISI:000278526700026} & JSSP   & {GA} & coop. multi-search & pC/RS/MPSS & randomly connected mesh & message passing & n/a \\
\citep{ISI:000301155300035} & {FSSP} & {GA} & low-level & 1C/RS/SPSS & master-slave & GPGPU & [16.6-19.1] \\
\hdashline
\citep{ISI:000358469500006} & AP    & {GA} & coop. multi-search & pC/RS/MPDS & grid  & message passing & n/a \\
\citep{ISI:000307760400022} & {FSSP} & {GA} & coop. multi-search & pC/RS/MPSS & randomly connected mesh & message passing & n/a \\
\citep{Homberger2008} & Other & {GA} & 1. coop. multi-search, 2. indep. multi-search & 1. pC/RS/MPSS, 2. pC/RS/MPSS & fully connected mesh & LAN file system & algorithm 1. (with migration): superlinear (n=30), algorithm 2. (independent): n/a \\
\citep{ISI:000266928200016} & MSP   & {GA} & coop. multi-search & pC/RS/MPSS & master-slave & message passing & n/a \\
\citep{ISI:000321889500001} & AP    & {GA} & coop. multi-search & pC/RS/MPSS & master-slave & message passing & linear (n=26-201) \\
\hdashline
\citep{ISI:000385241600006} & Other & {GA} & coop. multi-search & pC/RS/MPDS & master-slave & message passing & n/a \\
\citep{ISI:000345097100005} & MSP   & {GA} & coop. multi-search & pC/RS/SPDS & n/a   & n/a   & n/a \\
\citep{ISI:000388617500004} & {Other} & {GA} & coop. multi-search & pC/RS/MPDS & {master-slave} & {n/a} & superlinear (n=4-8) \\
\citep{ISI:000387234200020} & TSP   & {GA} & low-level & 1C/RS/SPSS & master-slave & GPGPU & n/a \\
\citep{ISI:000271409100010} & Other & {GA} & coop. multi-search & pC/RS/MPDS & ring  & n/a   & n/a \\
\hdashline
\citep{ISI:000398718400001} & BFP   & {GA} & {1. coop. multi-search, 2. low-level} & 1. pC/RS/MPSS 2. 1C/RS/SPSS & hybrid (ring + master-slave) & 1. shared memory, 2. hybrid (message passing + GPGPU) & n/a \\
\citep{ISI:000324107400005} & {GTP} & {GA} & {hybrid (low-level + coop. multi-search)} & 1C/RS/SPSS & master-slave & n/a   & varies between instances (n=24) \\
\citep{ISI:000312839600013} & Other & {GA} & coop. multi-search & pC/RS/MPDS & n/a   & n/a   & linear (n=2-8) \\
\citep{ISI:000312839600013} & Other & {PSO} & {n/a} & n/a & n/a   & n/a   & linear (n=2-8) \\
\citep{ISI:000300080700009} & Other & {PSO} & cooperative multi-search & 1. pC/RS/MPSS & master-slave & message passing & n/a \\
\hdashline
\citep{rao2017solving} & {GTP} & {PSO} & low-level & 1C/RS/SPSS & master-slave & GPGPU & n/a \\
\citep{ISI:000407295800001} & {GTP} & {PSO} & low-level & 1C/RS/SPSS & master-slave & GPGPU & n/a \\
\citep{ISI:000302315500003} & Other & {PSO} & low-level & 1C/RS/SPSS & communication via memory & GPGPU & [<=281.7] \\
\citep{ISI:000272541300004} & BFP   & {PSO} & 1. low-level, 2. low-level, 3. n/a & 1. 1C/RS/SPSS, 2. 1C/RS/SPSS, 3. n/a & 1. and 2. master-slave, 3. communication via memory & GPGPU & [1-28] \\
\citep{ISI:000293548900017} & Other & {PSO} & 1. domain decomposition, 2. coop. multi-search & 1. 1C/RS/MP?S, 2. pC/?/? & 1. communication via memory, 2. master-slave & GPGPU & [<45] \\
\citep{ISI:000401889300020} & Other & {SSPR} & {1. low-level, 2. indep. multi-search } & 1. 1C/RS/SPSS, 2. pC/RS/MPDS & n/a   & n/a   & n/a \\
\citep{bozejko2009solving} & FSSP   & {SSPR} & 1. low-level, 2. indep. multi-search & 1. 1C/RS/SPSS, 2. pC/RS/?  & master-slave & message passing & linear/superlinear (n=2-16) \\
\citep{Banos2014} & VRP   & {OEA} & coop. multi-search & pc/C/MPDS & master-slave & 1. message passing, 2. shared memory, 3. hybrid (message passing + shared memory) & 1. sublinear (n=4), 2. sublinear (m=4), 3 sublinear (n=2 nodes with m=2 threads each) \\
\hdashline
\citep{Deep2010} & BFP   & PSO   & coop. multi-search & 1. pC/RS/MPSS, 2. pC/RS/MPSS, 3. pC/RS/MPSS & master-slave & message passing & (type and development of slope over n) of speedup substantially varies between  benchmark functions (n=2-12) \\
\citep{Diego2012} & VRP   & ACO   & n/a   & n/a & {n/a} & GPGPU & [<13] \\
\citep{Ding2013} & BFP   & FA    & coop. multi-search & pC/RS/MPDS & {n/a} & GPGPU & [59.2- ~190] \\
\citep{Ding2013} & BFP   & PSO   & coop. multi-search & pC/RS/MPDS & {n/a} & GPGPU & [59.2- ~190] \\
\citep{Dongdong2010} & AP    & ACO   & 1. low-level, 2. indep. multi-search & 1. 1C/RS/SPSS, 2. pC/RS/MPSS & {1. master-slave, 2. n/a} & 1. shared memory, 2. threads & 1. linear/sublinear (n=1-32), 2. sublinear (n=1-32) \\
\hdashline
\citep{Lancinskas2013} & FLP   & GA    & coop. multi-search & pC/RS/MPDS & master-slave & message passing & linear/sublinear (n=128-1024) \\
\citep{Lancinskas2012} & FLP   & GA    & coop. multi-search & pC/RS/MPDS & master-slave & 1. message passing, 2. hybrid (message passing + shared memory)  & 1. linear/sublinear (n=32-512), sublinear (n=1024-2048), 2. sublinear (n= 32-512 nodes with m=4 threads each), sublinear (n= 8-128 nodes with m=16 threads each)  \\
\citep{Lazarova2008} & TSP   & {ACO} & 1. coop. multi-search, 2. ?, 3. coop. multi-search & 1. pC/?/MPDS, 2. ?, 3. pC/RS/MPDS & master-slave & hybrid (message passing + shared memory) & results vary between instances (n=2-10) \\
\citep{Lazarova2008} & TSP   & {GA} & 1. coop. mutli-search, 2. ?, 3. coop. multi-search & 1. pC/?/MPDS, 2. ?, 3. pC/RS/MPDS & ring  & hybrid (message passing + shared memory) & sublinear (n=2-10) \\
\citep{Nebro2010} & BFP   & OEA   & coop. multi-search & pC/RS/MPDS & {n/a} & threads & machine 1: linear (n=1-2), sublinear (n=4-32), machine 2: linear (n=1-32) \\
\citep{Nowotniak2011} & KP    & {OEA} & coop. multi-search & pC/RS/MPDS & {n/a} & GPGPU & [120 - 400] \\
\citep{Redondo2008} & FLP   & {OEA} & coop. multi-search & pC/RS/MPDS & 1. ring, 2. master-slave, 3. hybrid (ring + master-slave) & message passing & 1. linear (n=2-8), sublinear (n=16-32), 2. n/a, 3. n/a., 4. linear/super-linear (n=2), linear (n=4-8), linear/sub-linear (n=16), sub-linear (n=32)  \\
\citep{Sanci2011} & Other & GA    & 1.low-level, 2. cooperative multi-search & 1. 1C/RS/SPSS, 2. pC/RS/MPDS & {n/a} & GPGPU & [16.3-20.1] \\
\citep{Tsutsui2008} & AP    & {ACO} & {1. low-level, 2. indep. multi-search} & 1. 1C/RS/SPSS; 2. pC/C/? & 1.master-slave, 2a. fully connected mesh, 2b. other (replace-worst), 2c. other (unconnected) & threads & 1. sublinear (n=4; further, unspecified n), 2. sublinear (n=4) \\
\citep{Umbarkar2014} & BFP   & GA    & low-level & 1C/RS/SPSS & {fully connected mesh} & threads & n/a \\
\hdashline
\citep{Wang2008} & FLP   & {PSO} & domain decomposition & pC/KC/MPSS & master-slave & shared memory & sublinear (n=2) \\
\citep{Wang2012} & Other & {GA} & low-level & 1C/RS/SPSS & master-slave & GPGPU & [23-32.86] \\
\citep{Weber2011} & n/a   & {OEA} & coop. multi-search & pC/RS/MPDS & n/a   & n/a   & n/a \\
\citep{You2009} & TSP   & {ACO} & indep. multi-search & pC/?/MP?S & n/a   & GPGPU & [<22] \\
\citep{Zhao2011} & TSP   & GA    & coop. multi-search & pC/?/MPDS & n/a   & GPGPU & [3.3-5.3] \\
\hdashline
\citep{Zhao2011} & TSP   & ACO   & coop. multi-search & pC/?/MPDS & n/a   & GPGPU & [2.9-8.4] \\
\citep{Zhao2011} & TSP   & OEA   & coop. multi-search & pC/?/MPDS & n/a   & GPGPU & [3.15-15.8]\\
\citep{davidovic2011mpi} & MSP   & BCO   & 1. low-level, 2. indep. multi-search, 3. indep. multi-search & 1.1C/RS/SPSS, 2. pC/RS/SPDS, 3. pC/RS/SPDS & message passing   & MPI & 1. sublinear (n=2-12), 2. linear (n=2-5), 3. superlinear (n=2-12)\\
\citep{subotic2011different} & BFP   & BCO   & 1. indep. multi-search, 2. coop. multi-search, 3. coop. multi-search & 1. pC/RS/?SDS, 2. pC/RS/?SDS, 3. pC/RS/?SDS & threads & Java & 1. sublinear (n=4), 2. n/a, 3. n/a\\
\citep{izzo2009parallel} & BFP   & OEA   & coop. multi-search & pC/C/MPDS & threads & C++, POSIX threads & n/a\\
\hdashline
\citep{VALLADA2009365} & FSSP   & GA   & coop. multi-search & pC/C/?PDS & message passing & Delphi 2006, Msgconnect & n/a\\
\citep{ISI:000415593600024} & BFP   & GA & coop. multi-search & pC/RS/MP?S & master-slave & n/a   & n/a \\

 	\label{tab:results_populationBasedHeur}%
\end{longtable}%
\end{landscape}
\normalsize

\tiny
\begin{landscape}
\begin{longtable}{|p{3.5cm}p{1.2cm}p{1.2cm}p{2.8cm}p{2.3cm}p{1.7cm}p{2.5cm}p{4.2cm}|}
\hline
\multirow{2}{*}{\textbf{Reference}} & \multirow{2}{*}{\textbf{Problem}} & \multirow{2}{*}{\textbf{Algorithm}} & {\textbf{Parallelization}} & {\textbf{Process \&}} & {\textbf{Communication}} & {\textbf{Programming }} & \multirow{2}{*}{\textbf{Scalability}}\\
& & & {\textbf{strategy}} & {\textbf{search control}} & {\textbf{topology}} & {\textbf{model}} &\\ 
\hline
\endhead
\hline
\endfoot
\hline\\
 \caption{\normalsize Computational parallelization studies (hybrid metaheuristics)}
\endlastfoot

\citep{ISI:000379511800009} & JSSP   & HM (a. ACO  +  local search, b. ACO  +  SA) & 1a. low-level, 1b. coop. multi-search, 2. coop. multi-search & 1a. 1C/RS/SPSS, 1b. pC/C/?PDS, 2. pC/C/SPDS & n/a   & shared memory & n/a \\
\citep{ISI:000311921300006} & TSP   & HM (ACO  +  local search) & low-level & 1C/RS/SPSS & master-slave & GPGPU & a. [0.17-8.03], b. [0.06-23.6] \\
\citep{ISI:000344552400009} & Other & HM (OEA  +  local search) & 1. low-level, 2. low-level, 3. coop. multi-search & 1. 1C/RS/SPSS, 2. 1C/RS/SPSS, 3. pC/?/MP?S  & master-slave & 1. message passing, 2.shared memory, 3. hybrid (message passing + shared memory) & 1. linear (n=2-8), sublinear (n=16-64), 2. linear (n=2-8), 3. sublinear (n=16-64)  \\
\citep{ISI:000370099700013} & KP    & HM (OEA  +  local search) & low-level & 1C/RS/SPSS & {master-slave} & shared memory & varies between instances (n=24) \\
\citep{ISI:000298631400006} & MSP   & HM (OEA  +  local search) & coop. multi-search & pC/RS/MPDS & ring  & hybrid (shared memory + message passing) & n/a \\
\hdashline
\citep{ISI:000296012700016} & Other & HM (GA  +  local search) & low-level & 1C/RS/SPSS & master-slave & threads & sublinear/linear (n=2-16) \\
\citep{ISI:000295018500007} & MSP   & HM (GA  +  scheduling heuristic) & coop. multi-search & pC/C/MPDS & master-slave & n/a   & sublinear (n=8-64) \\
\citep{ISI:000285413400014} & Other & HM (GA  +  SA) & coop. multi-search & pC/RS/MPDS & tree  & n/a   & n/a \\
\citep{ISI:000405549600001} & TSP   & HM (GA  +  SA) & low-level & 1C/RS/SPSS & {master-slave} & {GPGPU} & n/a \\
\citep{ISI:000296539700061} & Other & HM (GA  +  TS) & coop. multi-search & pC/RS/MPDS & ring  & message passing & n/a \\
\hdashline
\citep{ISI:000271404200004} & Other & HM (GA  +  hill climbing) & hybrid (low-level + coop. multi-search) & n/a & grid  & {GPGPU} & [16-25] \\
\citep{ISI:000308548500016} & FSSP   & HM (GA  +  local search  +  greedy algorithms) & cooperative multi-search & pC/C/MPDS & fully-connected mesh & shared memory & n/a \\
\citep{ISI:000264988500031} & {GTP} & HM (GA  +  TS) & indep. multi-search & pC/RS/MPDS & master-slave & message passing & linear/sublinear (n=2-24) \\
\citep{Subramanian2010} & {VRP} & {HM (VNS  +  iterated local search)} & low-level & 1C/RS/SPSS & {master-slave} & {message passing} & sublinear (n=2-256) \\
\citep{ISI:000300080700009} & Other & HM (PSO + gradient descent) & cooperative multi-search &  pC/RS/MPSS & master-slave & message passing & n/a \\
\citep{ISI:000284747000034} & FSSP   & HM (cluster algorithm for simulated annealing + GA) & domain decomposition & 1C/RS/MPDS & master-slave & message passing & 1. linear (n=2-64), 2. n/a \\
\hdashline
\citep{ISI:000314737600028} & VRP   & HM (GA + SA) & coop. multi-search & pC/RS/MPDS & master-slave & message passing & sublinear (n=2-4) \\
\citep{ISI:000281591300145} & BFP   & HM (SA + differential evolution) & low-level & 1C/RS/SPSS & master-slave & message passing & linear (n=2-8) \\
\citep{Fujimoto2011} & TSP   & HM (OEA + 2-opt local search) & low-level & 1C/RS/SPSS & {n/a} & GPGPU & [9.3-24.2] \\
\citep{Ibri2010} & Other & HM (ACO + TS) & 1. indep. multi-search, 2. domain decomposition & 1. pC/C/?, 2. 1C/RS/? & master-slave & threads & sublinear (n=2-30) \\
\citep{Lancinskas2013} & FLP   & {HM (LS + GA)} & coop. multi-search & pC/RS/MPDS & master-slave & hybrid (message passing +shared memory)  & linear/sublinear (n=128-1024) (only algorithm version 2) \\
\hdashline
\citep{Taillard2012} & AP    & HM (algorithms unspecified) & domain decomposition & pC/RS/MPSS & n/a   & 1. threads, 2. GPGPU, 3. hybrid (threads + GPGPU) & 1. sublinear (n=4-8), 2. [7.0-14.4], 3. [9.8-16.7] \\
\citep{Xhafa2008} & MSP   & HM (algorithms unspecified) & 1. indep. multi-search, 2. low.level, 3. n/a & 1. pC/?/MPDS, 2. 1C/RS/SPSS, 3. n/a & 1. other (unconnected), 2. master-slave, 3. other (two-level: grid + mesh) & message passing & sublinear (n=3-9) \\
\citep{Zhao2011} & TSP   & {HM} & coop. multi-search & pC/?/MPDS & n/a   & GPGPU & [2.8-8.3] \\
\citep{Zhu2009} & BFP   & HM (ACO + pattern search) & low-level & 1C/RS/SPSS & master-slave & GPGPU & [66.85-403.91] 
 	\label{tab:results_hybridHeur}%
\end{longtable}%
\end{landscape}
\normalsize

\tiny
\begin{landscape}
\begin{longtable}{|p{3.5cm}p{1.2cm}p{1.2cm}p{2.8cm}p{2.3cm}p{1.7cm}p{2.5cm}p{4.2cm}|}
\hline
\multirow{2}{*}{\textbf{Reference}} & \multirow{2}{*}{\textbf{Problem}} & \multirow{2}{*}{\textbf{Algorithm}} & {\textbf{Parallelization}} & {\textbf{Process \&}} & {\textbf{Communication}} & {\textbf{Programming }} & \multirow{2}{*}{\textbf{Scalability}}\\
& & & {\textbf{strategy}} & {\textbf{search control}} & {\textbf{topology}} & {\textbf{model}} &\\ 
\hline
\endhead
\hline
\endfoot
\hline\\
 \caption{\normalsize Computational parallelization studies (problem-specific heuristics, other heuristics, matheuristics, and multi-search algorithms:)}
\endlastfoot

\citep{ISI:000355262900003} & FLP   & {MH (B-a-X  +  OEA)} & domain decomposition & pC/RS/MPSS & {n/a} & {shared memory} & {n/a} \\
\citep{ISI:000401878000008} & TSP   & {PSH} & domain decomposition & 1C/RS/MPSS & master-slave & message passing  & n/a \\
\citep{ISI:000290248600012} & VRP   & {MH (PSEA  +  OSSH)} & coop. multi-search & pC/KC/MPDS & master-slave & message passing & linear (n=8-64) \\
\citep{lahrichi2015integrative} & VRP & MS (integrative cooperative search method) & hybrid (domain decomposition and cooperative multi-search) & pC/KC/MPDS & communication via memory & shared memory & n/a \\
\citep{ISI:000299473600001} & AP & MS (several metaheuristics parallelized at the same time) & coop. multi-search & pC/RS/MPDS & master-slave & message passing & n/a\\
\hdashline
\citep{ISI:000418216800023} & MILP  & PSH   & coop. multi-search & pC/C/MPDS & master-slave & message passing & n/a \\
\citep{ISI:000386524100014} & {Other} & {PSH} & low-level & 1C/RS/SPSS & {n/a} & {threads} & linear/sublinear (n=2-8) \\
\citep{ISI:000349592500005} & {Other} & {PSH} & 1. domain decomposition, 2. coop. multi-search & 1. 1C/KS/SPSS; 2. pC/C/MPDS & {1. master-slave, 2. mesh} & {message passing} & 1. sublinear (n=4), 2. sublinear (n=7) \\
\citep{ISI:000321869500004} & {VRP} & {OH (Monte Carlo simulation inside a heuristic-randomization process)} & n/a   & n/a & n/a   & threads & n/a \\
\citep{ISI:000349884400003} & Other & PSH   & n/a   & n/a & master-slave & GPGPU & n/a \\
\hdashline
\citep{ISI:000349884400003} & Other & OH (method based on autonomous agents) & n/a   & n/a & master-slave & GPGPU & n/a \\
\citep{ISI:000255671800016} & Other & {PSH} & low-level & 1C/RS/SPSS & master-slave & message passing & n/a \\
\citep{ISI:000361986700006} & FLP   & {PSH} & low-level & 1C/RS/SPSS & master-slave\newline{} & 1. shared memory, 2. message passing & 1. linear (n=2-16), 2. linear (n=31-192) \\
\citep{ISI:000378100800023} & Other & {PSH} & domain decomposition & 1C/RS/MPSS & n/a   & GPGPU & [2-15] \\
\citep{bozejko2009solving} & FSSP   & PSH   & low-level & 1C/RS/SPSS & master-slave & n/a   & [2-3] (unreported n) \\
\citep{Dobrian2011} & GTP   & {PSH} & domain decomposition & pC/RS/MPSS & ring  & message passing & n/a \\
\citep{Ismail2011} & TSP   & {PSH} & {n/a} & n/a & master-slave & shared memory (task-oriented, implemented as threads) & sublinear (n=2)  \\
\hdashline
\citep{Luo2015} & Other & {PSH} & low-level & 1C/RS/SPSS & master-slave & threads & n/a \\
\citep{Sanci2011} & Other & {OH} & 1.low-level, 2. coop. multi-search & 1. 1C/RS/SPSS, 2. pC/RS/MPDS & {n/a} & GPGPU & [2.4-21.49] \\
\citep{Sathe2012} & GTP   & OH    & domain decomposition & pC/RS/MPSS & fully connected mesh & hybrid (message passing + shared memory) & relative speedup: sublinear/linear  (n=2-16), sublinear (n=32-1024); weak speedup: mixed results (n=1-1024)\\
\citep{ISI:000402046200007} & Other & MS (systolic neighborhood search + GA) & coop. multi-search & pC/RS/MPDS & master-slave & hybrid (shared memory + GPGPU) & n/a

 	\label{tab:results_prblSpecificOtherHeurMH}%
\end{longtable}%
\end{landscape}
\normalsize

\bibliographyappendix{x:/GUIDO/Literature/bibtex/Parallel_Optimization-CONSOLIDATED_WORKING_DOCUMENT(CWD),X:/GUIDO/Literature/bibtex/bhpc}
\normalsize

\end{document}